\documentclass[twocolumn,showpacs,preprintnumbers,amsmath,amssymb,aps,10pt,prb]{revtex4-1}
\usepackage{graphicx}

\newcommand{\me}[1]{\left< #1 \right>} 
\newcommand{\ket}[1]{\left| #1 \right>} 

\begin{document}

\title{Second harmonic generation spectroscopy of excitons in ZnO}

\author{M.~Lafrentz,$^{1}$ D.~Brunne,$^{1}$ A.~V.~Rodina,$^{2}$ V.~V.~Pavlov,$^{2}$
R.~V.~Pisarev,$^{2}$ D.~R.~Yakovlev,$^{1,2}$ A.~Bakin,$^{3}$ and M.~Bayer$^{1}$}
\affiliation{$^{1}$Experimentelle Physik 2,Technische Universit\"at Dortmund,
44221 Dortmund, Germany} \affiliation{$^{2}$Ioffe Physical-Technical Institute, Russian Academy of Sciences, 194021 St. Petersburg, Russia} \affiliation{$^{3}$Institut f\"ur Halbleitertechnik, Technische Universit\"at Braunschweig, 38106 Braunschweig, Germany}

\begin{abstract}
Nonlinear optics of semiconductors is an important field of
fundamental and applied research, but surprisingly the role of
excitons in the coherent processes leading to harmonics generation
has remained essentially unexplored. Here we report results of a
comprehensive experimental and theoretical study of the three-photon
process of optical second harmonic generation (SHG) involving the
exciton resonances of the noncentrosymmetric hexagonal wide-band-gap
semiconductor ZnO in the photon energy range of $3.2-3.5$~eV.
Resonant crystallographic SHG is observed for the $1s(A,B)$,
$2s(A,B)$, $2p(A,B)$, and $1s(C)$ excitons. We show that strong SHG
signals at these exciton resonances are induced by the application
of a magnetic field when the incident and the SHG light wave vectors
are along the crystal $z$-axis where the crystallographic SHG
response vanishes. A microscopic theory of SHG generation through
excitons is developed, which shows that the nonlinear interaction of
coherent light with excitons has to be considered beyond the
electric-dipole approximation. Depending on the particular symmetry
of the exciton states SHG can originate from the electric- and
magnetic-field-induced perturbations of the excitons due to the
Stark effect, the spin as well as orbital Zeeman effects, or the
magneto-Stark effect. The importance of each mechanism is analyzed
and discussed by confronting experimental data and theoretical
results for the dependencies of the SHG signals on photon energy,
magnetic field, electric field, crystal temperature, and light
polarization. Good agreement is obtained between experiment and
theory proving the validity of our approach to the complex problem
of nonlinear interaction of light with ZnO excitons. This general
approach can be applied also to other semiconductors.
\end{abstract}

\maketitle

\section{Introduction}

Nonlinear optics has opened multifaceted possibilities for studying
and tailoring light-matter interaction. Nowadays nonlinear optical
phenomena and materials are a broad basis for fundamental and
applied research \cite{Bloembergen,Shen,Boyd,Bass,Nikogosyan}. In
linear optics, propagation, absorption, and emission of light are
essentially single-photon processes. In contrast, in nonlinear
optics the interaction of light with a medium is governed by
multi-photon processes. Obviously the light-matter interaction
becomes more intricate thereby. Linear and nonlinear optical
experiments address different types of optical susceptibilities,
still they all are determined by the features of the crystal
structure of the material under study as well as the resulting
charge and spin properties. Therefore they open versatile
opportunities for in-depth analysis from different perspectives. In
this sense, linear and nonlinear optics may be regarded as
independent and simultaneously complementary to each other in
material investigations.

In nonlinear optics, frequency conversion processes such as second,
third, and higher order harmonics generation as well as the sum and
difference frequency generation play a particularly important role
\cite{Bloembergen,Shen,Boyd}. Among these phenomena, most prominent
is the simplest three-photon process $(\omega,\omega,-2\omega)$ of
optical second harmonic generation (SHG): first, the parity
selection rules for the optical transitions between the contributing
electronic states radically differ from those in linear optics and
also from other, more complicated nonlinear phenomena; second,
besides the parity selection rules, the time-reversal symmetry
operation is a principally important factor in harmonics generation
when the spin system becomes involved in applied magnetic fields or
in magnetically ordered materials
\cite{Fiebig1,Pisarev-pss,Pisarev-JL}.

From the beginning of nonlinear optics, optical generation of second
and higher order harmonics has been subject of active research in
various semiconductors \cite{Garmire,Garmire2}. However, the
majority of these studies, typically performed on bulk crystals or
thin films, were limited to fixed excitation wavelengths when the
fundamental and harmonics photon frequencies were in the
transparency region, below the fundamental band gap. This approach
is motivated by avoiding any absorption in the medium which would
impede potential applications. There are only few examples where the
SHG spectroscopic studies of semiconductors covered broad spectral
ranges. Absolute values of the SHG coefficients for bulk zinc-blende
ZnTe, ZnSe, and ZnS  were measured at room temperature in the SHG
spectral range $1.8-4.8$~eV \cite{Wagner}. Spectroscopic SHG in bulk
GaAs was reported in the range $2-5$~eV, covering several electronic
transitions at critical points \cite{Bergfeld}. The SHG spectral
features found for these materials were in reasonable agreement with
theoretical calculations and experimental data acquired by other
techniques. For hexagonal ZnO, the material selected for the present
study, SHG was reported for selected wavelengths below the band gap
in numerous publications, see e.g. \cite{Miller,Wang,Wang2,Abe}.
Further, also spectroscopic SHG studies in the range $2.25-3.44$~eV
were reported for ZnO microcrystallite thin films \cite{Zhang}. The
SHG output was found to increase significantly in the vicinity of
the direct band gap. SHG over broad infrared spectral ranges was
also applied to various chalcopyrites, semiconductors of practical
importance, see e.g. \cite{Medvedkin,Kumar} and references therein.

In semiconductors the optical properties in close vicinity of the
band gap are largely determined by excitons, bound complexes of an
electron and a hole. The exciton energy levels including their spin
properties have been intensely studied using both linear and
nonlinear optical methods such as absorption, reflection,
photoluminescence, two-photon absorption, four-wave mixing,
\emph{etc} \cite{Garmire,Garmire2}. Surprisingly, the contributions
of excitons to harmonics generation have remained essentially
unexplored. Typically studies lack a microscopic theoretical
explanation, with scarce exceptions \cite{Chang,Leitsmann,Pedersen}.
Experimental observations were reported for forbidden SHG in
resonance with the $2p$ Wannier exciton in ZnSe thin films
\cite{Minami,Minami1}; resonant SHG at the $1s$ orthoexciton level
in Cu$_2$O \cite{MYShen,Kono}; and second and third harmonic
spectroscopy of excitons in a homoepitaxial GaN layer
\cite{Schweitzer}. An early attempt to detect SHG signals in the
spectral region of the $C$ exciton in ZnO and the $1s$ excitons in
CuCl was undertaken in Refs.~\cite{Haueisen0,Haueisen}. The concept
of Wannier excitons was used in the SHG study of CuCl, whereas
Frenkel excitons were explored in SHG studies of a C$_{60}$
molecular crystal \cite{Janner1,Janner2}.

Detail insight into the role of excitons in harmonics generation can
be taken if the studies are performed at low temperatures with high
spectral resolution. External magnetic and electric fields can
perturb and mix charge and spin states, providing novel mechanisms
for nonlinear harmonics generation. For example, in diamagnetic
materials GaAs and CdTe the orbital quantization is the origin of
magnetic-field-induced SHG \cite{Pavlov1,Sanger2}. In diluted
magnetic semiconductors (Cd,Mn)Te, on the other hand, the giant
Zeeman spin splitting was shown to be the source of
magnetic-field-induced SHG \cite{Sanger1}. Even for the
centrosymmetric magnetic semiconductors EuTe and EuSe a magnetic
field was found to induce SHG with high efficiency
\cite{Kaminski,Kaminski1}.

This rudimentary state of the exciton SHG problem has motivated our
spectroscopic research of SHG as the simplest frequency conversion
process in the wide-band-gap semiconductor ZnO, characterized by a
large exciton binding energy of $60$~meV and a rich exciton level
structure \cite{Klingshirn1}. This material has recently gained
substantial renewed interest, partly because the large exciton
binding energy could lead to lasing by exciton recombination even at
room temperature. This and other potential ZnO applications are
discussed in the comprehensive review by \"{O}zg\"{u}r {\it et al}
\cite{Alivov}. In order to get deeper insight, our study of SHG at
excitons is performed in applied magnetic and electric fields. In
conjunction with a detailed theoretical analysis we show that SHG
spectroscopy allows us to work out the underlying microscopic
mechanisms of the nonlinear process of simultaneous coherent
two-photon excitation and subsequent one-photon emission involving
excitons. Magnetic and electric fields can perturb the exciton
states through the Stark, the magneto-Stark, and the Zeeman effects
and may act therefore as sources of SHG carrying characteristic
signatures for the chosen field geometry. Our findings open new
opportunities for studying exciton complexes in detail and involving
them in frequency conversion processes.

The paper is organized as follows. In Sec.~\ref{sec:symmetrySHG}, we
describe the crystallographic and electronic structures as well as
the optical and magneto-optical properties of hexagonal ZnO. Also a
symmetry analysis of the SHG polarization selection rules is given.
The details of the experiment are presented in
Sec.~\ref{sec:experiment}, followed by Sec.~\ref{sec:results} where
the experimental data are shown. In Sec.~\ref{sec:theory} the
microscopic theory of the SHG is introduced and several mechanisms
involving exciton states are suggested. The comparison of experiment
and theory in Sec.~\ref{sec:discussion} allows assignment of
particular signals in the SHG to a specific mechanism, which to out
knowledge has been mission so far. The developed understanding can
be applied also to other semiconductor materials.

\section{Second harmonic generation in Z\lowercase{n}O}
\label{sec:symmetrySHG}

\subsection{Symmetry of electronic states, excitons and polaritons}
\label{subsec:symmetry}

\begin{figure}
\includegraphics[width=0.45\textwidth,angle=0]{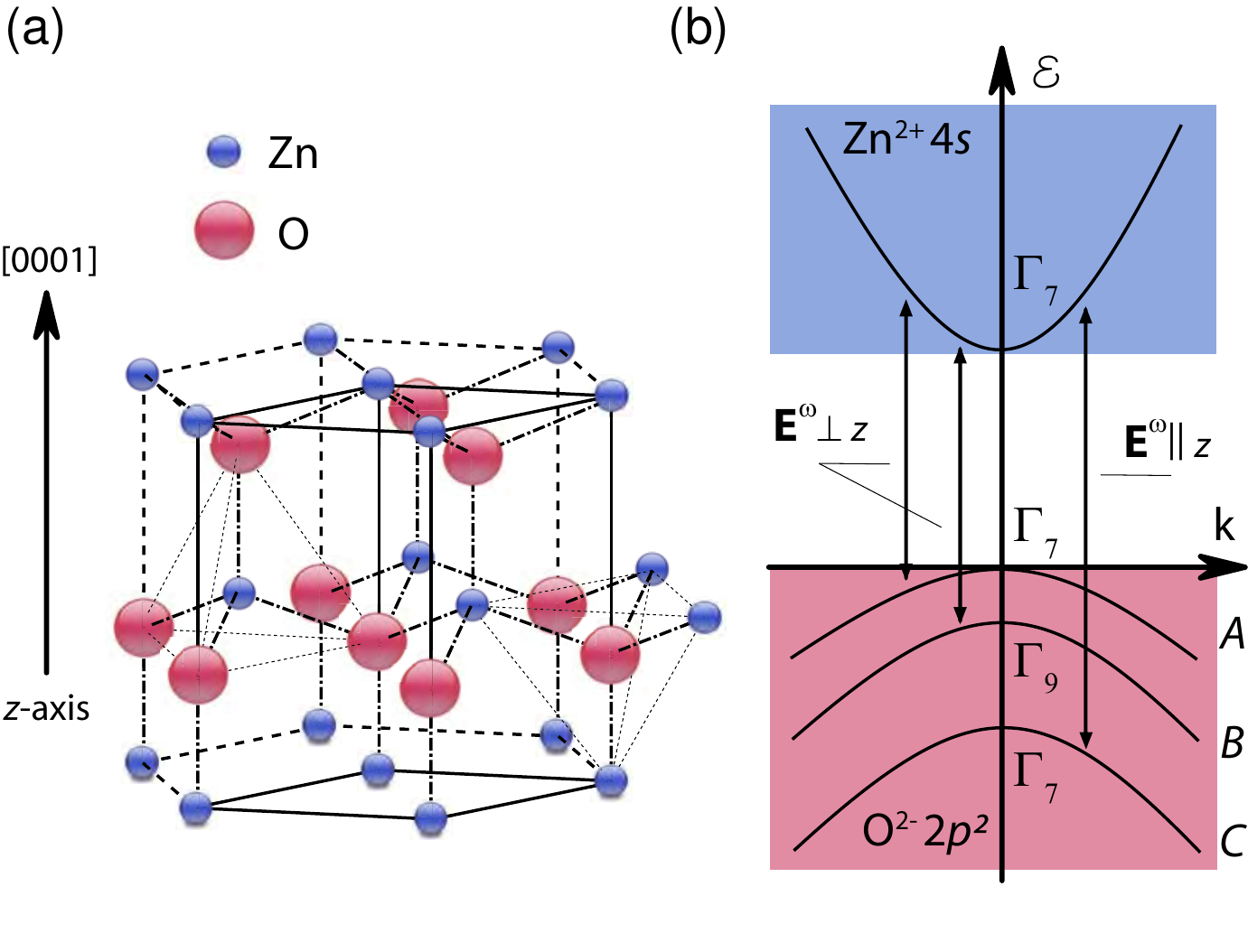}
\caption{(color online) (a) Uniaxial crystal structure of wurtzite
ZnO, $[0001]\parallel \mathbf{z}$ is the hexagonal crystallographic
axis. (b) Electronic band structure of wurtzite ZnO. The hexagonal
crystal field is responsible for the energy splitting between the
$A$, $B$, and $C$ valence bands. }\label{fig:figure1}
\end{figure}

ZnO crystallizes preferably in the wurtzite-type structure
\cite{LB,Alivov}, see Fig.~\ref{fig:figure1}(a), characterized by
two interconnected sublattices of Zn$^{2+}$ and O$^{2-}$ ions with a
strong ionic binding. The lattice constants of ZnO are $a_0 =
3.2495$ {\AA} and $c_0 = 5.2069$ {\AA} \cite{Jagadish}. The unit
cell is formed by two Zn$^{2+}$ and two O$^{2-}$ ions ($Z=2$) each
of them being tetrahedrally surrounded by four ions of the other
species. Wurtzite ZnO has a hexagonal crystal lattice, belonging to
the point group $6mm$ and space group P$6_3mc$
\cite{Hahn,Klingshirn3}. The $z$-direction of the used Cartesian
$xyz$-system is chosen parallel to the polar hexagonal
$[0001]$-axis, the so-called $c$-axis, which subsequently will be
referred to as $z$-axis following the Birss notation,
$[0001]\parallel \mathbf{z}$ and $[1\overline{1}00]\parallel
\mathbf{y}$ \cite{Birss}.  From the optical point of view ZnO is a
uniaxial material with the optical axis directed along the
crystallographic $z$-axis.

The electronic band structure of wurtzite ZnO is shown in
Fig.~\ref{fig:figure1}(b). The valence band is formed by the $2p^2$
orbitals of the O$^{2-}$ ions and the conduction band is formed by
the $4s$ orbitals of the Zn$^{2+}$ ions. The the $2p$-levels (and
the antibonding $sp^3$ orbitals) are split by the hexagonal crystal
field into two $\Gamma_5$ and $\Gamma_1$ subbands. Including the
spin through the spin-orbit interaction leads to a further splitting
into three twofold degenerate valence band states $(\Gamma_1 \oplus
\Gamma_5)\oplus \Gamma_7=\Gamma_7 \oplus \Gamma_9 \oplus \Gamma_7$.
In all wurtzite-type semiconductors these bands are usually labeled
from higher to lower energies as $A$ ($\Gamma_9$), $B$ ($\Gamma_7$)
and $C$ ($\Gamma_7$) bands. However, ZnO has an inverted valence
band ordering $A$ ($\Gamma_7$), $B$ ($\Gamma_9$) and $C$
($\Gamma_7$) \cite{Klingshirn1}. The selection rules for transitions
from the upper valence bands $A$ ($\Gamma_7$) and $B$ ($\Gamma_9$)
to the conduction band ($\Gamma_7$) are essentially the same,
because the admixture of the $\ket{z}$ character to the Bloch wave
functions of the $A$ ($\Gamma_7$) valence band is small
\cite{Lambrecht}. As a result, these transitions  are allowed for
$\mathbf{E}^{\omega} \perp \mathbf{z}$, where $\mathbf{E}^{\omega}$
is the electric field of the fundamental light wave. Transitions
from the $C$ ($\Gamma_7$) valence band to the conduction band
($\Gamma_7$) are allowed for $\mathbf{E}^{\omega}\parallel
\mathbf{z}$.

Correspondingly, three exciton series are formed in ZnO by a
$\Gamma_7$ electron and a hole from one of the $A$ ($\Gamma_7$), $B$
($\Gamma_9$), or $C$ ($\Gamma_7$) valence bands. These excitons have
approximately the same binding energy of $\simeq 60$~meV and a Bohr
radius of $\simeq 1.8$~nm. The exciton symmetry results from the
direct product of the envelope function symmetry and the symmetry of
conduction and valence band Bloch states, see e.g.,
Ref.~\cite{Wheeler}. The energies of the resulting exciton states
are split by the short-range exchange interaction. For the
$s$-symmetry excitons of the $A$ and $B$ series, the strongest state
has $\Gamma_5$ symmetry. It is twofold degenerate and polarized
perpendicular to the $z$-axis, while for the $C$ exciton it has
$\Gamma_1$ symmetry and is polarized parallel to the $z$-axis. As a
result, for light propagating along the $z$-axis ($\mathbf{k}
\parallel \mathbf{z}$) both $\Gamma_5$ excitons are transversal,
while the $\Gamma_1$ exciton is longitudinal and cannot be excited.
For light with $\mathbf{k} \perp \mathbf{z}$, one of the $\Gamma_5$
exciton states is transverse and the other is longitudinal, while
the $\Gamma_1$ exciton is transversal. The resonances of the
longitudinal excitons are shifted to higher energies by the
long-range exchange interaction.

The strong light-matter interaction in ZnO leads to the formation of
exciton-polaritons and their symmetries depend on the direction of
the light propagation. The interaction of the transverse excitons
with photons leads to the formation of two transverse, lower (LPB)
and upper (UPB), polariton branches. Their dispersion relations can
be obtained from the condition
$\varepsilon_\bot(\omega,\mathbf{k})=(kc/\omega)^2$ for the
$\Gamma_5$ excitons and
$\varepsilon_\|(\omega,\mathbf{k})=(kc/\omega)^2$ for the $\Gamma_1$
excitons. Here $c$ is the speed of light and $\omega$ is the photon
frequency. $\varepsilon_\bot$ and $\varepsilon_\|$ are the
dielectric functions for the electric field of light polarized
perpendicular and parallel to the $z$-axis, respectively, including
contributions of exciton resonances with energies close to $\hbar
\omega$. The energies of the UPB at $k=0$ coincide with the energies
of the longitudinal excitons determined from
$\varepsilon_{\bot,\|}(\omega,k=0)=0$, while the energies of the LPB
at $k\rightarrow\infty$ coincide with the energies of the
transversal excitons. If $\mathbf{k}$ is not parallel or
perpendicular to the crystal $z$-axis, one obtains the so called
mixed-mode polaritons \cite{Ivchenko,Wrzesinski,Klingshirn2}.

\subsection{Polarization selection rules for SHG}
\label{subsec:selectionrules}

Wurtzite ZnO belongs to the noncentrosymmetric point group $6mm$
and, consequently, the leading-order SHG is allowed in the
electric-dipole (ED) approximation. The crystallographic SHG
polarization $P^{2\omega}$ can be written as
\begin{equation}
P_{i}^{2\omega}= \epsilon_0
\chi^{\text{cryst}}_{ijl}(\omega,\omega,-2\omega)E_j^\omega E_l^\omega
 , \label{eq:P1}
\end{equation}
where $i,j,l$ are the Cartesian indices,  $\epsilon_0$ is the vacuum
permittivity, $\chi^{\text{cryst}}_{ijl}$ is the nonlinear optical
susceptibility, $E^{\omega}_{j(l)}$ are the components of the
fundamental light electric field $\mathbf{E}^{\omega}$. In the ED
approximation and in absence of external fields a group theoretical
analysis predicts the following nonzero components of the
crystallographic nonlinear optical susceptibility for bulk ZnO
$\chi^{\text{cryst}}_{ijl}$ :
$\chi_{xxz}=\chi_{xzx}=\chi_{yyz}=\chi_{yzy}$,
$\chi_{zxx}=\chi_{zyy}$, and $\chi_{zzz}$ \cite{Boyd,Popov}.

Note, that Eq.~\eqref{eq:P1} accounts only for the ED contributions  on- or off-resonant with electronic
band transitions at the fundamental and SHG photon frequencies
$\omega$ and 2$\omega$. However, more generally the SHG process can
involve also electric-quadrupole (EQ) and magnetic-dipole (MD)
contributions. They become important when the outgoing SHG is
resonant with the exciton energy ${\cal E}_{\mathrm{exc}}$, for
example. Taking into account higher order contributions and the
feasibility of a resonance, the incoming fundamental electric field
$\mathbf{E}^\omega(\mathbf{r},t) = \mathbf{E}^\omega
\exp[{\mathrm{i}(\mathbf{k}\mathbf{r} -\omega t)}]$ generates an
effective polarization inside the semiconductor at the double
frequency as \cite{Sionnest88}:
\begin{eqnarray}
P_{\mathrm{eff},i}^{2\omega}({\cal E}_{\mathrm{exc}})=\epsilon_0
\chi^{\text{cryst}}_{ijl}({\cal
E}_{\mathrm{exc}},\mathbf{k}_{\mathrm{exc}}) E_j^\omega E_l^\omega
\, , \label{eq:P2}
\end{eqnarray}
where the nonlinear optical susceptibility
$\chi^{\text{cryst}}_{ijl}({\cal
E}_{\mathrm{exc}},\mathbf{k}_{\mathrm{exc}})$ describes the
spatial-dispersion phenomena entering in the EQ and MD
approximation. $\mathbf{k}_{\mathrm{exc}}=2n\mathbf{k}$ is the
exciton wave vector, $n$ is the refractive index at the fundamental
energy $\hbar \omega$, and $\mathbf{k}$ is the wave vector of the
incoming light.

Additional information on the exciton energy levels including their
spin structure, as well as on their wave functions can be obtained
by applying external fields. The symmetries of exciton states may be
modified by electric or magnetic fields, enabling mixing of states.
This opens the way for novel SHG mechanisms induced by the fields.
In this case, the effective polarization inside the semiconductor
can be written as
\begin{equation}
P_{\mathrm{eff,B,E},i}^{2\omega}({\cal E}_{\mathrm{exc}})=
\epsilon_0 \chi_{ijl}({\cal
E}_{\mathrm{exc}},\mathbf{k}_{\mathrm{exc}},\mathbf{B},\mathbf{E})
E_j^\omega E_l^\omega \, , \label{eq:P3}
\end{equation}
where the nonlinear optical susceptibility $\chi_{ijl}({\cal
E}_{\mathrm{exc}},\mathbf{k}_{\mathrm{exc}},\mathbf{B},\mathbf{E})$
accounts for phenomena induced by the external magnetic
($\mathbf{B}$) and electric ($\mathbf{E}$) fields. The nonlinear
polarizations in Eqs.~\eqref{eq:P1}-\eqref{eq:P3} are the sources of
the outgoing SHG electric field  $\mathbf{E}^{2\omega}(\mathbf{r},t)
\propto \mathbf{P}^{2\omega}\exp[{\mathrm{i}(2\mathbf{k}\mathbf{r}
-2\omega t)}]$ with SHG intensity $I^{2 \omega}\propto
|\mathbf{P}^{2\omega}|^2$.

For a resonant SHG process, which involves the ground state of the
unexcited crystal $|G\rangle$ and an exciton state
$|\text{Exc}\rangle$, the optical transition from $|G\rangle$ to
$|\text{Exc}\rangle$ should be allowed both for
the two-photon excitation and the one-photon emission process.
Fulfillment of this condition strongly depends on the crystal
symmetry and experimental geometry. With including excitonic
effects, this situation becomes richer due to the different
symmetries of the exciton states with $s$, $p$ and $d$ type of
envelope wave functions.

Further, external perturbations such as stress,  electric or
magnetic field can mix the exciton states, thereby reducing their
symmetry. For linear optical spectroscopy on ZnO (e.g. one-photon
absorption or emission) only  $s$ exciton states are active, while
$p$ states cannot be seen. In order to study $p$ exciton states,
either nonlinear spectroscopy (e.g. two-photon absorption) or
external perturbations, which mix $s$ and $p$ states  have to
be used. To study exciton states and their mixing, we performed
detailed experimental and theoretical studies of ZnO, serving as a
model system, by SHG spectroscopy with application of magnetic and
electric fields.

\section{Experiment}
\label{sec:experiment}

\begin{figure}
\includegraphics[width=0.4\textwidth,angle=0]{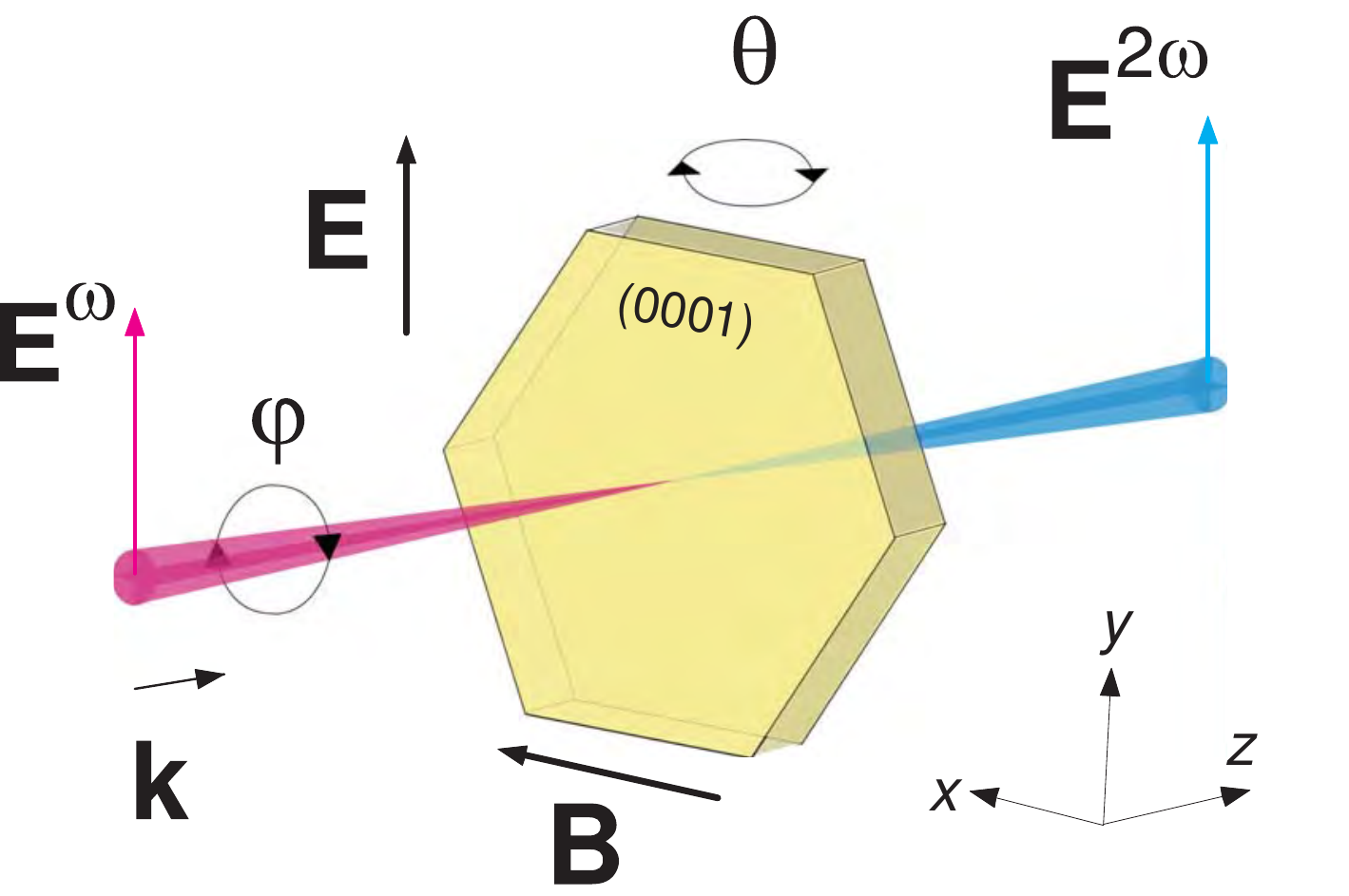}
\caption{(color online) Sketch demonstrating the measurement
geometry. $\theta$ is the sample tilting angle, $\varphi$ is the
turning angle of $\mathbf{E}^{\omega(2\omega)}$ around $\mathbf{k}$.
Electric and magnetic fields are perpendicular to each other and to
the propagation direction of the light
$\mathbf{E}\perp\mathbf{B}\perp\mathbf{k}$.} \label{fig:figure2}
\end{figure}

A hydrothermally grown hexagonal ZnO crystal of high optical quality
with $[0001]$ orientation and thickness of 500~$\mu$m was chosen for
this study. The SHG technique used for exciton spectroscopy was
described in Ref.~\cite{Sanger2}. The linearly polarized fundamental
light with photon energy $\hbar\omega$ was provided by a laser
system with an optical parametric oscillator tunable in the spectral
range of interest ($\hbar\omega=1.6-1.75$~eV) and generating optical
pulses of $7$~ns duration with energies up to $3$~mJ per pulse. The
experiments were performed in the transmission geometry with the
light wave vector $\mathbf{k}$ either parallel or tilted to the
$z$-axis of the ZnO sample. The SHG signal at photon energies
$2\hbar\omega$ was spectrally selected by a monochromator and
detected by a cooled charge-coupled-device camera. The experimental
geometry is shown in Fig.~\ref{fig:figure2}. Here $\theta$ is the
angle between the light wave vector $\mathbf{k}$ and the $z$-axis
which gives the sample tilting. $\varphi$ is the azimuthal angle of
the fundamental light polarization, where $\varphi =0^{\circ}$
coincides with the crystallographic $y$-axis. Magnetic fields
$\mathbf{B}$ up to $10$~T generated by a split-coil superconducting
solenoid were applied in the Voigt geometry ($\mathbf{B}\perp
\mathbf{k}$) or the Faraday geometry ($\mathbf{B}\parallel
\mathbf{k}$). External electric fields $\mathbf{E}$ up to $550$~V/cm
were applied via contacts perpendicular both to the magnetic field
and the propagation direction of the light
$\mathbf{E}\perp\mathbf{B}\perp\mathbf{k}$. The sample temperature
$T$ was varied in the range $1.6-125$~K.

As we will show below in Secs.~\ref{sec:results} and
Sec.~\ref{sec:theory}, decisive experiments for distinguishing
different microscopic contributions to the SHG signal are
measurements of the rotational anisotropy, i.e., the dependence of
the SHG signal on the azimuthal angle. Such rotational anisotropies
were measured for four different geometries:

1.
$I^{2\omega}_\parallel\mapsto\mathbf{E}^{2\omega}\parallel\mathbf{E}^{\omega}$,
fundamental and SHG light polarizations are rotated synchronously
such that they are parallel to each other.

2.
$I^{2\omega}_\perp\mapsto\mathbf{E}^{2\omega}\perp\mathbf{E}^{\omega}$,
fundamental and SHG light polarization are rotated synchronously,
such that the SHG light polarization is perpendicular to the
fundamental light.

3.
$I^{2\omega}_{\parallel\mathbf{B}}\mapsto\mathbf{E}^{2\omega}\parallel\mathbf{B}$,
SHG light polarization is fixed parallel to the magnetic field
direction while the fundamental light polarization is rotated around
$\mathbf{k}$.

4.
$I^{2\omega}_{\perp\mathbf{B}}\mapsto\mathbf{E}^{2\omega}\perp\mathbf{B}$,
SHG light polarization is fixed perpendicular to the magnetic field
direction while the fundamental light polarization is turned around
$\mathbf{k}$.

The corresponding patterns of rotational anisotropies are modeled
according to Eqs.~(1-3). The results are discussed in
Sec.~\ref{sec:discussion}.

\section{Experimental results}
\label{sec:results}

\subsection{Crystallographic SHG}
\label{subsec:cryst_SHG}

\begin{figure}
\includegraphics[width=0.45\textwidth,angle=0]{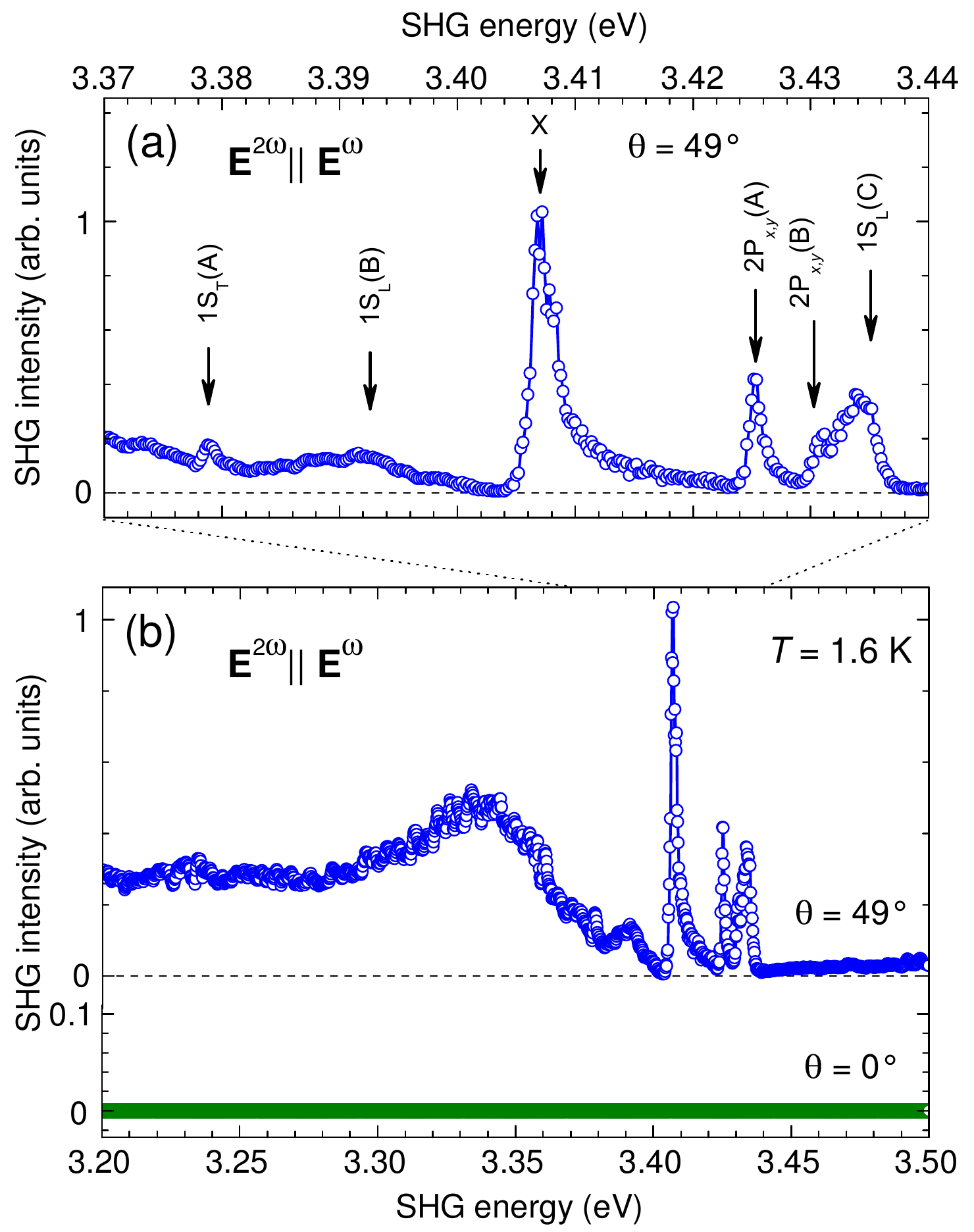}
\caption{(color online) Crystallographic SHG spectra of ZnO for
$\mathbf{E}^{2\omega}\parallel \mathbf{E}^\omega$ and $\varphi =
90^\circ$ (compare with anisotropies in Fig.~\ref{fig:figure4}),
measured at $T=1.6$~K. (a) Close-up of the exciton spectral range
$3.37-3.44$~eV for $\theta=49^\circ$. (b) Extended spectral range
$3.2-3.5$~eV for $\theta=49^\circ$ and $0^\circ$. SHG signals vanish
for $\mathbf{k}\parallel \mathbf{z}$ ($\theta = 0^\circ$).}
\label{fig:figure3}
\end{figure}

It follows from the symmetry analysis of the selection rules in
Sec.~\ref{subsec:selectionrules}, that for laser light propagating
along the hexagonal $z$-axis ($\mathbf{k}\parallel \mathbf{z}$) no
ED crystallographic SHG is allowed. This geometry addresses solely
components of the susceptibility without $z$ index, which are all
zero. Indeed, no SHG signals are found experimentally in the
vicinity of the ZnO band gap for zero tilting angle
$\theta=0^\circ$, see Fig.~\ref{fig:figure3}(b).

\begin{figure}
\includegraphics[width=0.45\textwidth,angle=0]{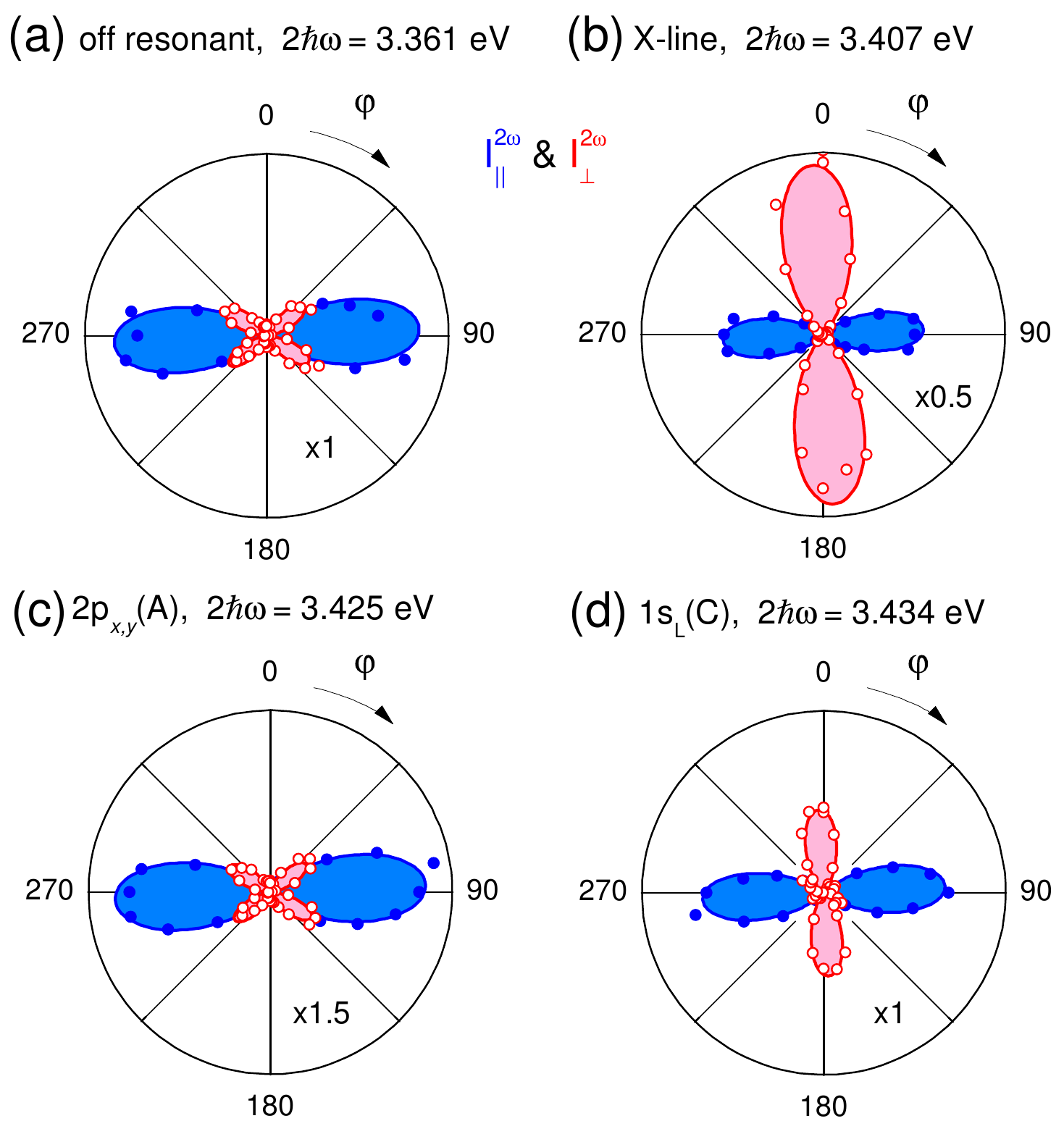}
\caption{(color online) Angular dependencies of the crystallographic
SHG measured at $\theta=49^\circ$ for different energies. Filled
(blue) and  open (red) circles represent the geometries
$\mathbf{E}^{2\omega}\parallel \mathbf{E}^\omega$ and
$\mathbf{E}^{2\omega}\perp \mathbf{E}^\omega$, respectively. Lines
and shaded areas show best fits according to Eq.~(1). (a) Off
resonant signal at $2\hbar\omega=3.361$~eV. (b) Unidentified \textit{X}-line
at $2\hbar\omega=3.407$~eV. Signal is scaled down by a factor of
$2$. (c) $2p(A)$ exciton line at $2\hbar\omega=3.425$~eV. Signal is
scaled up by a factor of $1.5$. (d) $1s_{\text{L}}$(C) exciton line
at $2\hbar\omega=3.434$~eV.} \label{fig:figure4}
\end{figure}

For tilted geometry, $\mathbf{k} \angle \mathbf{z}\neq0$, SHG is
provided by the nonzero $\chi^{\text{cryst}}_{ijl}$ components with
$z$ index, see Eq.~\eqref{eq:P1}. For parallel orientation of the
linear polarizations of fundamental and SHG light
($\mathbf{E}^{2\omega}\parallel \mathbf{E}^\omega$), strong SHG
signals are found for $\varphi = 90^\circ$ (see
Fig.~\ref{fig:figure4}). As one can see in
Fig.~\ref{fig:figure3}(b), the SHG consists of a broad band in the
spectral range below the exciton transitions. Further, several sharp
lines show up in the exciton spectral range. Above the band gap the
SHG signal vanishes. The SHG intensity shows pronounced rotational
anisotropies summarized in Fig.~\ref{fig:figure4}. These
anisotropies allow separation of the SHG signals from two-photon
photoluminescence signals which are expected to be isotropic.

The exciton spectral range is shown in more detail in
Fig.~\ref{fig:figure3}(a). Arrows mark the the reported energies of
the different exciton states in hexagonal ZnO \cite{Fiebig2}.
Letters T and L mark the transversal and longitudinal excitons,
respectively. A feature at $3.407$~eV has not been reported in
literature so far. It will be referred to as the $X$-line and
discussed in more detail in Sec.\ref{sec:discussion}. In the tilted
geometry, both $s$ and $p$ exciton states are SHG-active. Using
linear spectroscopy, only the $1s$ exciton states are observed in
absorption and reflection spectra due to their strong oscillator
strength. By contrast, the SHG intensity in the range of $1s$ states
of $A$ and $B$ excitons is surprisingly much weaker than the
intensity of their excited $2s$ and $2p$ states. This observation is
a clear manifestation that nonlinear SHG spectroscopy addresses
exciton properties inaccessible by linear spectroscopy.

Rotational anisotropy diagrams were measured for
$\mathbf{E}^{2\omega}\parallel \mathbf{E}^\omega$ and
$\mathbf{E}^{2\omega}\perp \mathbf{E}^\omega$ at the following
spectral positions: $3.361$, $3.379$, $3.391$, $3.400$, $3.407$,
$3.413$, $3.425$, $3.430$, $3.434$, and $3.444$~eV. Those
anisotropies in Figs.~\ref{fig:figure4}(a) and ~\ref{fig:figure4}(c)
are representative for the off-resonant and $A, B$ exciton regions.
Figs.~\ref{fig:figure4}(b) and \ref{fig:figure4}(d) show rotational
anisotropies specific for energies close to the $X$-line and the
$1s_{\text{L}}(C)$ exciton. The strongest signal for
$\mathbf{E}^{2\omega}\parallel \mathbf{E}^\omega$ is found for all
energies in the ($yz$)-plane meaning that the shape is dominated by
the $\chi_{zzz}$ component. Indeed, the fitting procedure gives an
$\chi_{zzz}$ value which is an order of magnitude larger than the
other components. For the crossed geometry
$\mathbf{E}^{2\omega}\perp \mathbf{E}^\omega$, the fitting procedure
gives the same ratio of nonlinear components. On the other hand, for
energies close to the $1s_{\text{L}}(C)$ exciton and the $X$-line in
Figs.~\ref{fig:figure4}(b) and \ref{fig:figure4}(d), the SHG
intensity $I^{2\omega}_\perp$ has a pronounced feature along the
$y$-axis, which can be explained by a phase shift. The real and
imaginary parts of the nonlinear components change signs in these
specific regions, leading to a strong distortion in the crossed
geometry $\mathbf{E}^{2\omega}\perp \mathbf{E}^\omega$.

We note, that our experiments do not allow us to measure the absolute
values of the nonlinear susceptibilities. Therefore, it is difficult
to compare the relative nonlinearities for SHG signals which are
widely separated on the photon energy scale. To measure the absolute
values of the nonlinear susceptibilities one has to take into
account the complex linear refraction indices for both the
fundamental and the SHG photon energies \cite{Haueisen}. We also
note, that the expected quadratic increase of the SHG intensity with
the fundamental power has been confirmed for the off-resonant and
on-resonant crystallographic SHG signals, as well as for the
magnetic-field-induced SHG signals, described in the next
subsection.

\subsection{Magnetic-field-induced SHG}
\label{subsec:MFI_SHG}

\begin{figure}
\includegraphics[width=0.45\textwidth,angle=0]{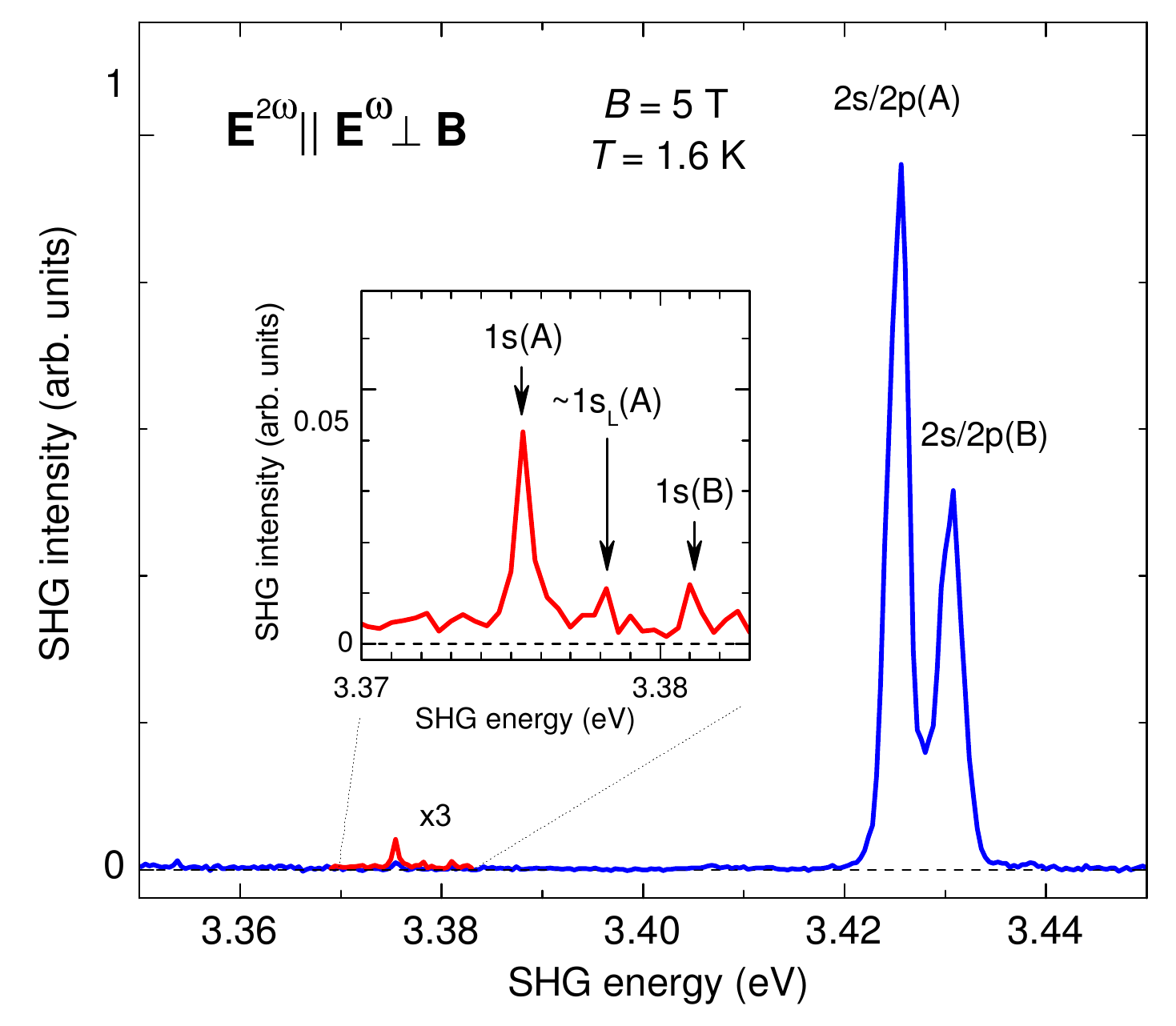}
\caption{(color online) Magnetic-field-induced SHG spectrum of ZnO
in a wide energy range $3.35-3.45$~eV for
$\mathbf{E}^{2\omega}\parallel \mathbf{E}^\omega\perp \mathbf{B}$
and $\varphi = 0^\circ$ at $T=1.6$~K. Inset shows the $1s$ exciton
region zoomed by a factor of $20$. The integration time for
recording the data shown by the red line was tripled compared to the
data shown by the blue line.} \label{fig:figure5}
\end{figure}

For investigating magnetic-field-induced SHG (MFISH)  we chose the
experimental geometry with $\theta=0^\circ$, where the
crystallographic SHG signal vanishes. A magnetic field $B=5$~T was
applied perpendicular to the $z$-axis in the geometry
$\mathbf{B}\perp \mathbf{k}\parallel \mathbf{z}$. SHG signals are
observed only in resonance with excitons, see
Fig.~\ref{fig:figure5}. Two strong lines are seen in the spectral
range of the $2s/2p(A,B)$ exciton states and three much weaker lines
in the range of the $1s(A,B)$ excitons. Note, that this behavior
differs from the observations in GaAs and CdTe, where the $1s$
exciton line always dominates in the MFISH spectrum
\cite{Pavlov1,Sanger2}. In ZnO, the weak SHG intensity observed at
the $1s(A,B)$ excitons can be related to the strong absorption of
SHG light due to the large absolute values of the complex dielectric
function \cite{Cobet}.

A magnetic field is an axial vector of even parity and, therefore,
it is not supposed to mix wave functions of opposite parities. In
spite of this restriction, a strong magnetic-field-induced
contribution to SHG was found in ZnO. An in-depth analysis of the
SHG microscopic mechanisms is required for understanding these
experimental findings. Such analysis based on the theoretical model
of Sec.~\ref{sec:theory} will be given in Sec.~\ref{sec:discussion}.

\subsubsection{SHG on $1s(A,B)$ excitons}
\label{subsubsec:1s_SHG}

\begin{figure}
\includegraphics[width=0.45\textwidth,angle=0]{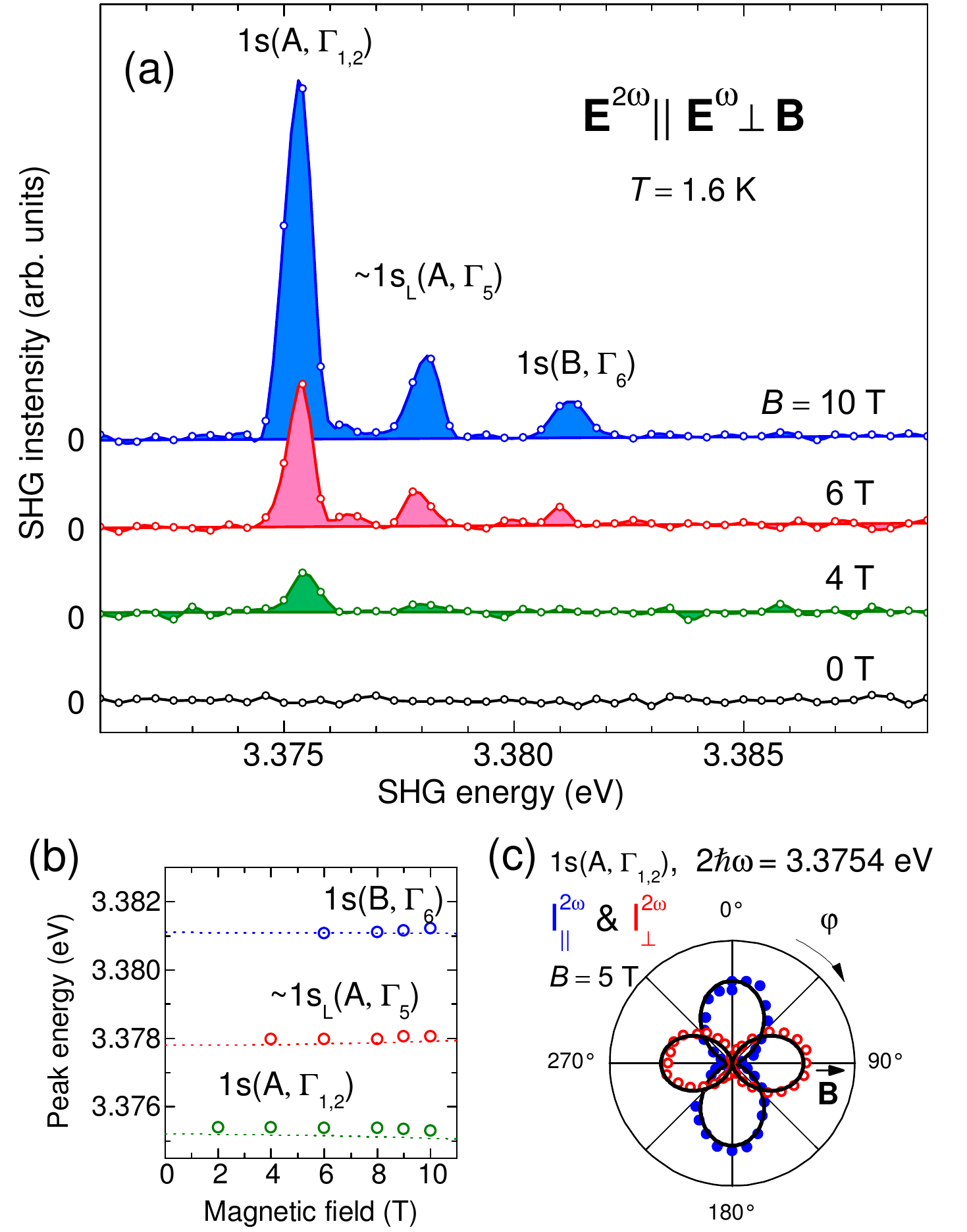}
\caption{(color online) (a) Magnetic-field-induced SHG spectra in
the range of the $1s(A,B)$ excitons in ZnO. (b) Magnetic field
dependence of the peak energy of the SHG exciton lines from panel
(a), dashed lines are calculated after Eq.~\eqref{eq:S8}. (c) Rotational
anisotropy of SHG intensity measured for the strongest line at
$B=5$~T, detected for synchronous rotation of the linear polarizers
for fundamental and SHG light: blue filled circles for
$\mathbf{E}^{2\omega}\parallel \mathbf{E}^{\omega}$  and red open
circles for $\mathbf{E}^{2\omega} \perp \mathbf{E}^{\omega}$. Black
lines give best fits after Eqs.~\eqref{eq:anieq4} and
\eqref{eq:anieq6}.} \label{fig:figure6}
\end{figure}

\begin{figure}
\includegraphics[width=0.45\textwidth,angle=0]{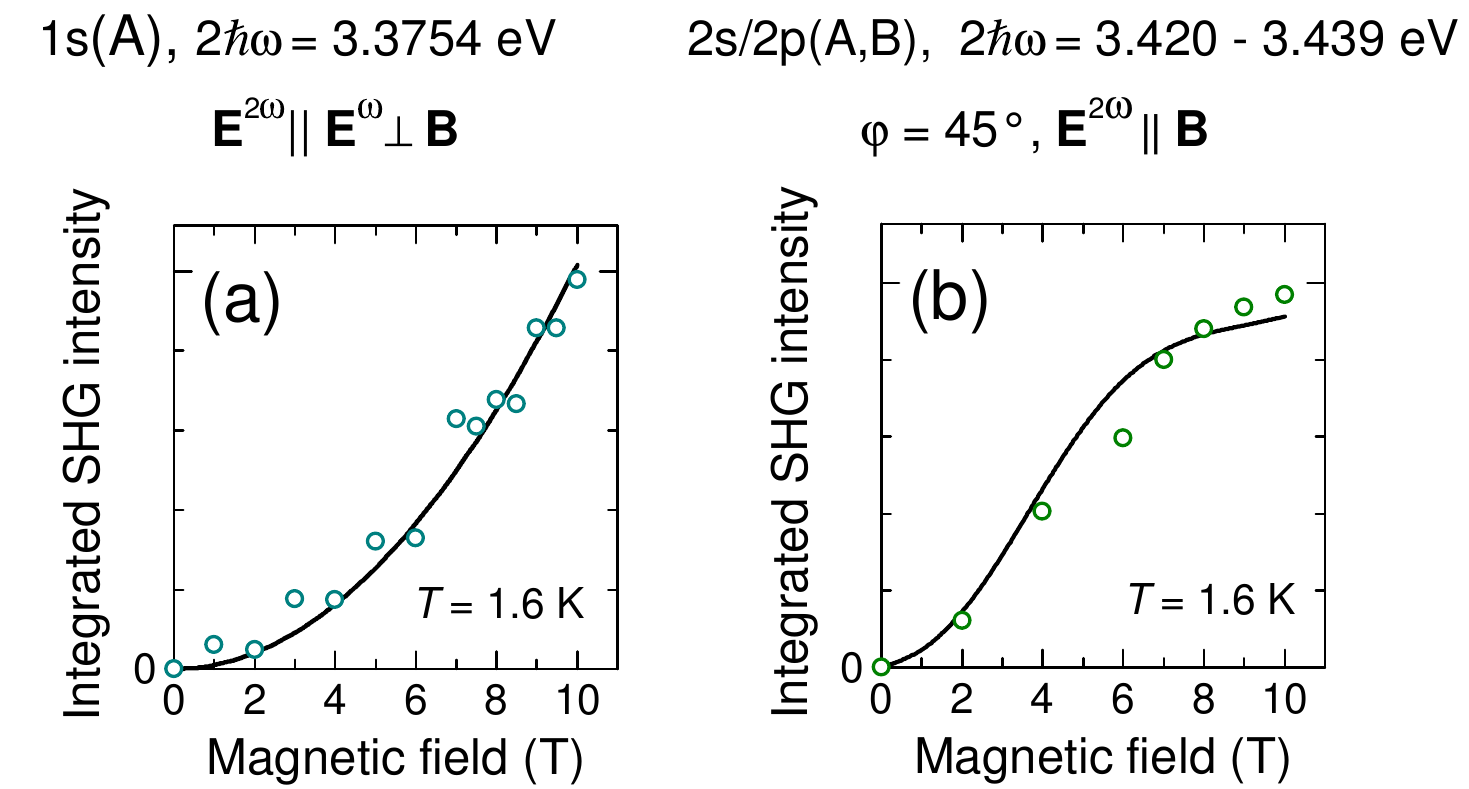}
\caption{(color online) (a) Integrated SHG intensity for the
strongest $1s$ line at $2\hbar\omega=3.3754$~eV [compare
Fig.~\ref{fig:figure6}(a)] as function of magnetic field (symbols).
Line is the best fit with $I^{2\omega}_\parallel\propto B^2$. (b)
Integrated SHG intensity in the spectral range of the $2s/2p(A,B)$
excitons at $2\hbar\omega=3.420-3.439$~eV (compare
Fig.~\ref{fig:figure8}) as function of magnetic field (symbols).
Line gives model calculation for $2\hbar\omega=3.4254$~eV (the
energy of the strongest peak in the SHG spectra) and
$\Gamma^i=1.2$~meV.}\label{fig:figure7}
\end{figure}

Let us consider the experimental observations in the $1s$ exciton
region in magnetic field more thoroughly. The corresponding SHG
spectra for different fields up to $10$~T are shown in
Fig.~\ref{fig:figure6}(a). Three lines corresponding to the $1s(A)$
paraexciton, the $1s(A,B)$ middle polariton branch \cite{comment},
and the $1s(B)$ paraexciton are clearly seen at the strongest field
with the first line at $3.3754$~eV being the most intense. The
integrated intensity of the $1s(A,\Gamma_{1,2})$ line shows a $B^2$
dependence, see Fig.~\ref{fig:figure7}(a). Its temperature
dependence measured at $B=7$~T shows a rapid decrease; see
Fig.~\ref{fig:figure14}(a). For excitons with strong binding
energies it is expected that the diamagnetic shift of their $1s$
states for the magnetic field strengths used here is very small, as
illustrated in Fig.~\ref{fig:figure6}(b).

Figure~\ref{fig:figure6}(c) shows the rotational anisotropies of the
SHG intensities in  a magnetic field of $5$~T for the parallel
$\mathbf{E}^{2\omega}\parallel\mathbf{E}^{\omega}$ and the
perpendicular $\mathbf{E}^{2\omega}\perp \mathbf{E}^{\omega}$
detection geometries. The rotational anisotropies show twofold
symmetry patterns, have the same amplitudes and are rotated relative
to each other by $90^{\circ}$, where the strongest signal is found
for $\varphi=0^\circ$ in the parallel geometry. These patterns
clearly differ from the crystallographic ones in
Fig.~\ref{fig:figure4}, highlighting the difference of involved SHG
mechanisms. The fits are done according to Eqs.~\eqref{eq:anieq4}
and \eqref{eq:anieq6} in Sec.~\ref{sec:theory}.

\subsubsection{SHG at $2s/2p(A,B)$ excitons}
\label{subsubsec:2s2p_SHG}

\begin{figure}
\includegraphics[width=0.45\textwidth,angle=0]{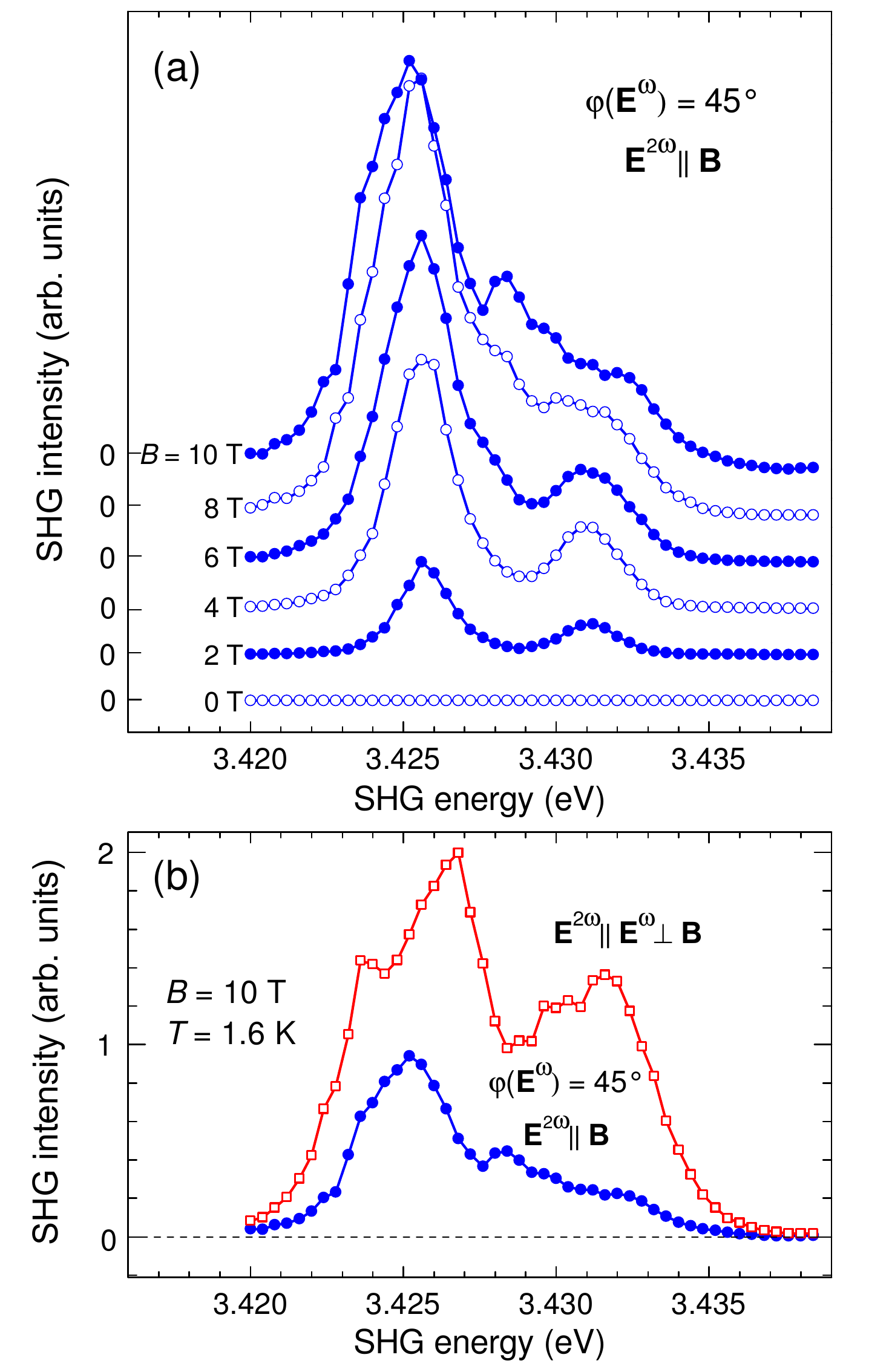}
\caption{(color online) Magnetic-field-induced SHG spectra in the
energy range of the $2s/2p$ exciton states measured at $T=1.6$~K.
(a) SHG spectra in different magnetic fields for
$\mathbf{E}^{2\omega}\parallel \mathbf{B}$ and
$\varphi(\mathbf{E}^{\omega}) = 45^\circ$. (b) SHG spectra at
$B=10$~T for $\mathbf{E}^{2\omega} \parallel \mathbf{E}^{\omega}
\perp \mathbf{B}$ and $\mathbf{E}^{2\omega}\parallel \mathbf{B}$
with $\varphi(\mathbf{E}^{\omega}) = 45^\circ$. }\label{fig:figure8}
\end{figure}

\begin{figure}
\includegraphics[width=0.45\textwidth,angle=0]{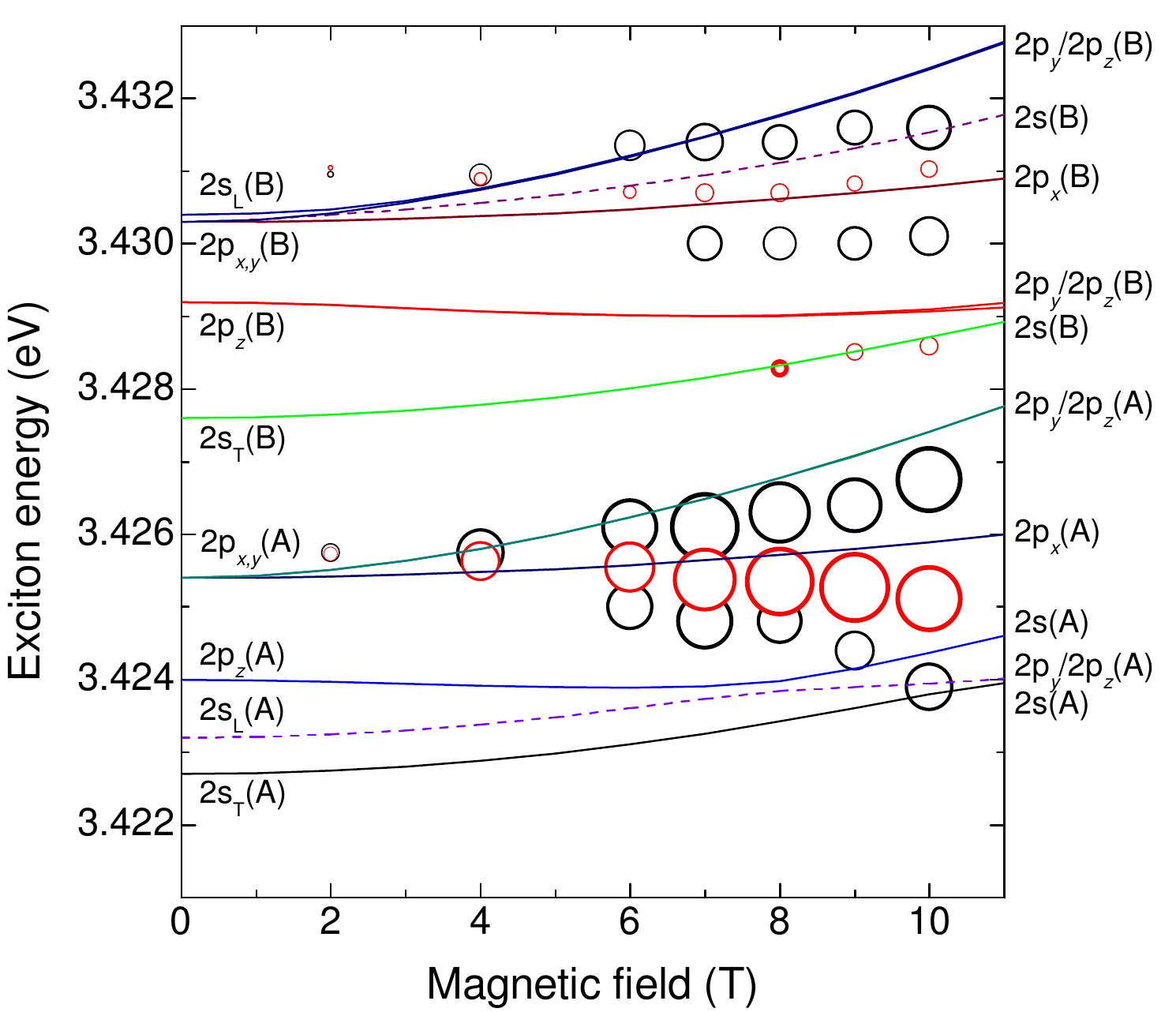}
\caption{(color online) Fan chart diagram for the magnetic field
dependencies of the $2s$ and $2p$ excited state energies of the $A$
and $B$ excitons. Symbols are experimental data with their size
scaled by the observed peak intensity. Black and red circles are
observed in the geometries $\mathbf{E}^{2\omega} \parallel
\mathbf{E}^{\omega} \perp \mathbf{B}$ and $\mathbf{E}^{2\omega}
\parallel \mathbf{B}$ with $\varphi(\mathbf{E}^{\omega})=45^\circ$,
respectively. Solid lines give the energies of the coupled
$2s/2p_z/2p_y$ and the $2p_x$ states according to the calculations
in Sec.~\ref{sec:Appendix} and after Eq.~\eqref{eq:S12}. Labels at
low fields indicate the zero field exciton energies, while labels at
high fields give the dominant component in the mixed exciton wave
functions. }\label{fig:figure9}
\end{figure}

Figure~\ref{fig:figure8}(a) shows the magnetic-field-induced SHG
spectra at the $2s/2p(A,B)$ exciton photon energies. A double peak
structure with lines at $3.425$ and $3.431$~eV appears with
increasing magnetic field, corresponding to the energies of the
$2p(A,B)$ excitons \cite{Dinges1,Fiebig3}. In strong magnetic fields
exceeding $7$~T the doublet structure splits further into at least
four peaks. In fact, more states can be distinguished in high
magnetic fields when the signals in different polarization
geometries are analyzed, see Fig.~\ref{fig:figure8}(b). The energy
shifts of these lines are plotted as a fan chart diagram in
Fig.~\ref{fig:figure9}, where the SHG peak intensities are
represented by the symbol sizes.

The magnetic field dependence of the integrated SHG intensity of the
$2s/2p$ states shows a quadratic increase at low fields $B<6$~T,
similar to the $1s$ exciton, and then tends to saturate for $B>6$~T,
see Fig.~\ref{fig:figure7}(b). This behavior, which was not observed
for the $1s$ states, gives a strong hint that the mechanisms
responsible for the magnetic-field-induced SHG differ for the $1s$
and the $2s/2p$ excitons in ZnO. The temperature dependence of the
$2s/2p$ SHG intensity in Fig.~\ref{fig:figure14}(a) shows a similar
but slightly faster decrease than that of the $1s$ excitons.

\begin{figure}
\includegraphics[width=0.45\textwidth,angle=0]{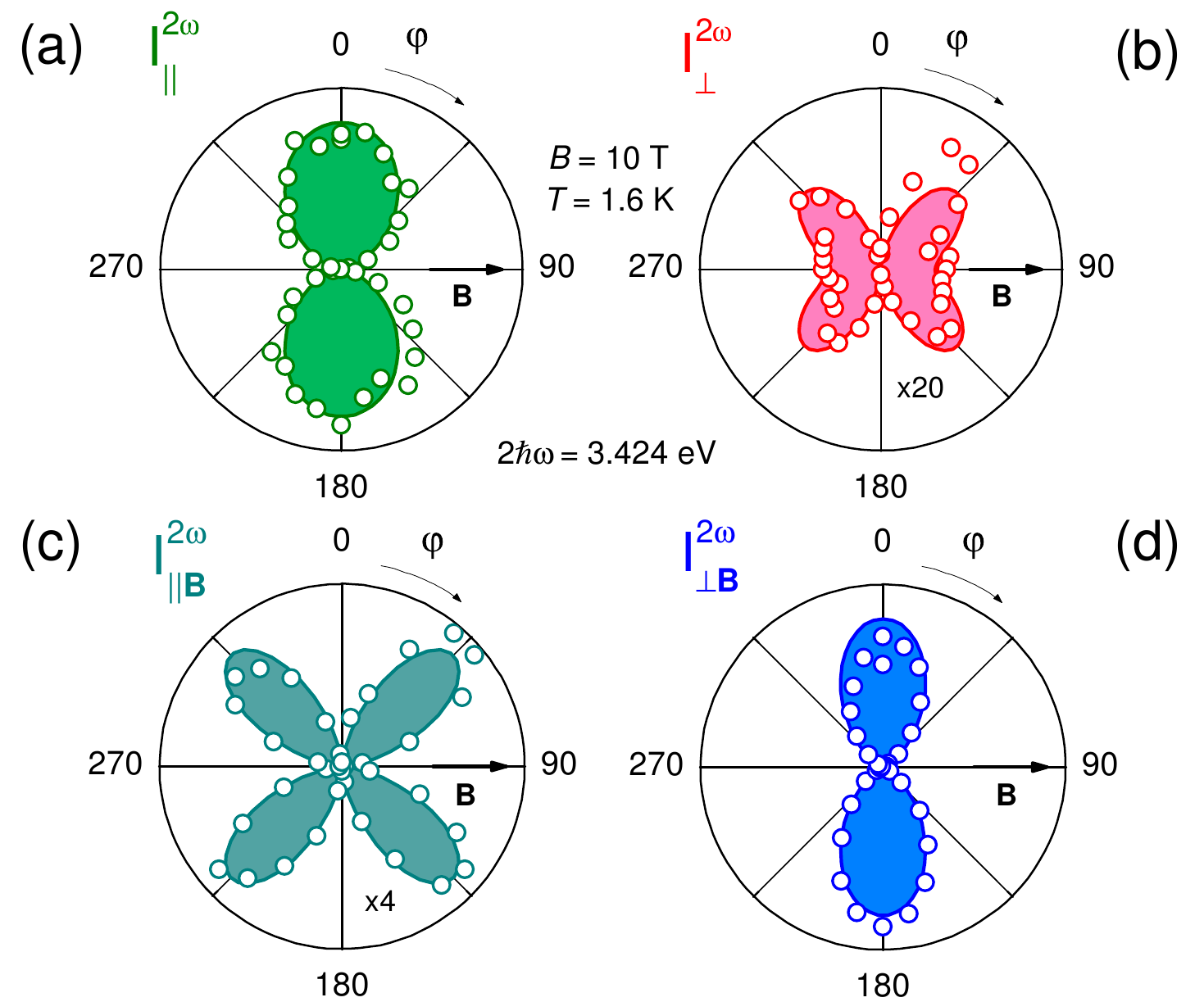}
\caption{ Angular dependencies of the magnetic-field-induced SHG
intensity at $3.424$~eV for different geometries at $B=10$~T. Open
circles represent measured data and lines show best fits following
Eqs.~\eqref{eq:anieq1}-\eqref{eq:anieq6}. (a)
$\mathbf{E}^{2\omega}\parallel \mathbf{E}^\omega$; fit according to
Eq.~\eqref{eq:anieq4}. (b) $\mathbf{E}^{2\omega}\perp
\mathbf{E}^\omega$; fit according to
\\$I^{2\omega}\propto\left[a\sin\varphi+b\sin\varphi\cos^2\varphi\right]^2$
with $a/b=1/2$; $a$ and $b$ represent the spin Zeeman
contributions of '$s$'- and '$p$'-type, respectively, see Figs.~\ref{fig:figure16}(b) and \ref{fig:figure16}(j).
(c) $\mathbf{E}^{2\omega}\parallel \mathbf{B}$; fit
according to Eq.~\eqref{eq:anieq2}. (d) $\mathbf{E}^{2\omega}\perp \mathbf{B}$;
fit according to $I^{2\omega}\propto\cos^4\varphi$.
}\label{fig:figure10}
\end{figure}

\begin{figure}
\includegraphics[width=0.45\textwidth,angle=0]{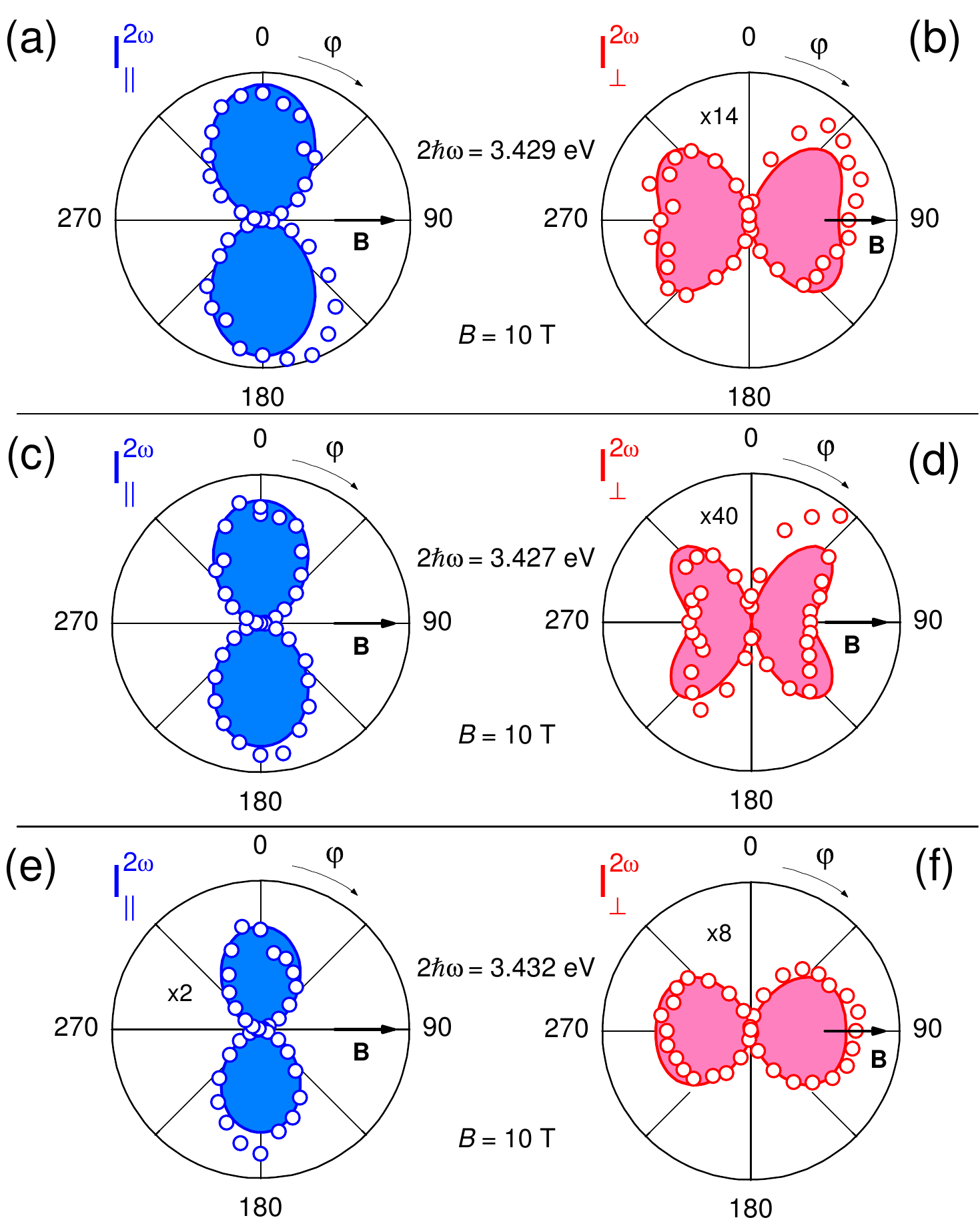}
\caption{Angular dependencies of the magnetic-field-induced SHG intensity for
different energies at $B=10$~T. Open blue circles give the measured
intensity $I_\parallel^{2\omega}$ for $\mathbf{E}^{2\omega}\parallel
\mathbf{E}^\omega$ and open red circles give the measured intensity
$I_\perp^{2\omega}$ for $\mathbf{E}^{2\omega}\perp
\mathbf{E}^\omega$. Solid lines show best fits according to
$I_\parallel^{2\omega}\propto \chi_{yyy}^2\cos^2(\varphi)$ (see
Eq.~\eqref{eq:anieq4}) and
$I^{2\omega}_\perp\propto\left[a\sin\varphi+b\sin\varphi\cos^2\varphi\right]^2$;
$a$ and $b$ represent the spin Zeeman contributions of '$s$'- and
'$p$'-type excitons, respectively, see Figs.~\ref{fig:figure16}(b)
and \ref{fig:figure16}(j). The ratio $I_\parallel^{2\omega}/I^{2\omega}_\perp$
indicates the dominance of the magneto-Stark contribution compared
to the spin Zeeman contribution. \\ (a, b) $2\hbar\omega=3.429$~eV;
$I_\parallel^{2\omega}/I^{2\omega}_\perp\approx 14/1$;
$I^{2\omega}_\perp$ with $a/b=1/1$. \\ (c, d) $2\hbar\omega=3.427$~eV;
$I_\parallel^{2\omega}/I^{2\omega}_\perp\approx 40/1$;
$I^{2\omega}_\perp$ with $a/b=3/4$. \\ (e, f) $2\hbar\omega=3.432$~eV;
$I_\parallel^{2\omega}/I^{2\omega}_\perp\approx 3/1$;
$I^{2\omega}_\perp$ with $a/b=2/1$. }\label{fig:figure11}
\end{figure}

SHG rotational anisotropies detected in different geometries at
$B=10$~T for the $2s/2p(A,B)$ states are shown in
Figs.~\ref{fig:figure10} and \ref{fig:figure11}. They have different
patterns depending on the exciton state involved, see, e.g.,
Figs.~\ref{fig:figure10}(b) and \ref{fig:figure11}(c). The
anisotropies differ strongly in the amplitude ratios for different
geometries. In the perpendicular geometry $\mathbf{E}^{2\omega}
\perp \mathbf{E}^{\omega}$ the shapes are quite different, compare
Figs.~\ref{fig:figure11}(b) and \ref{fig:figure11}(f), leading to
the assumption that the responsible SHG mechanisms are different and
vary with the photon energy.

\subsubsection{Magnetic-field-induced versus crystallographic SHG}
\label{subsubsec:B_vs_cryst_SHG}

\begin{figure}
\includegraphics[width=0.45\textwidth,angle=0]{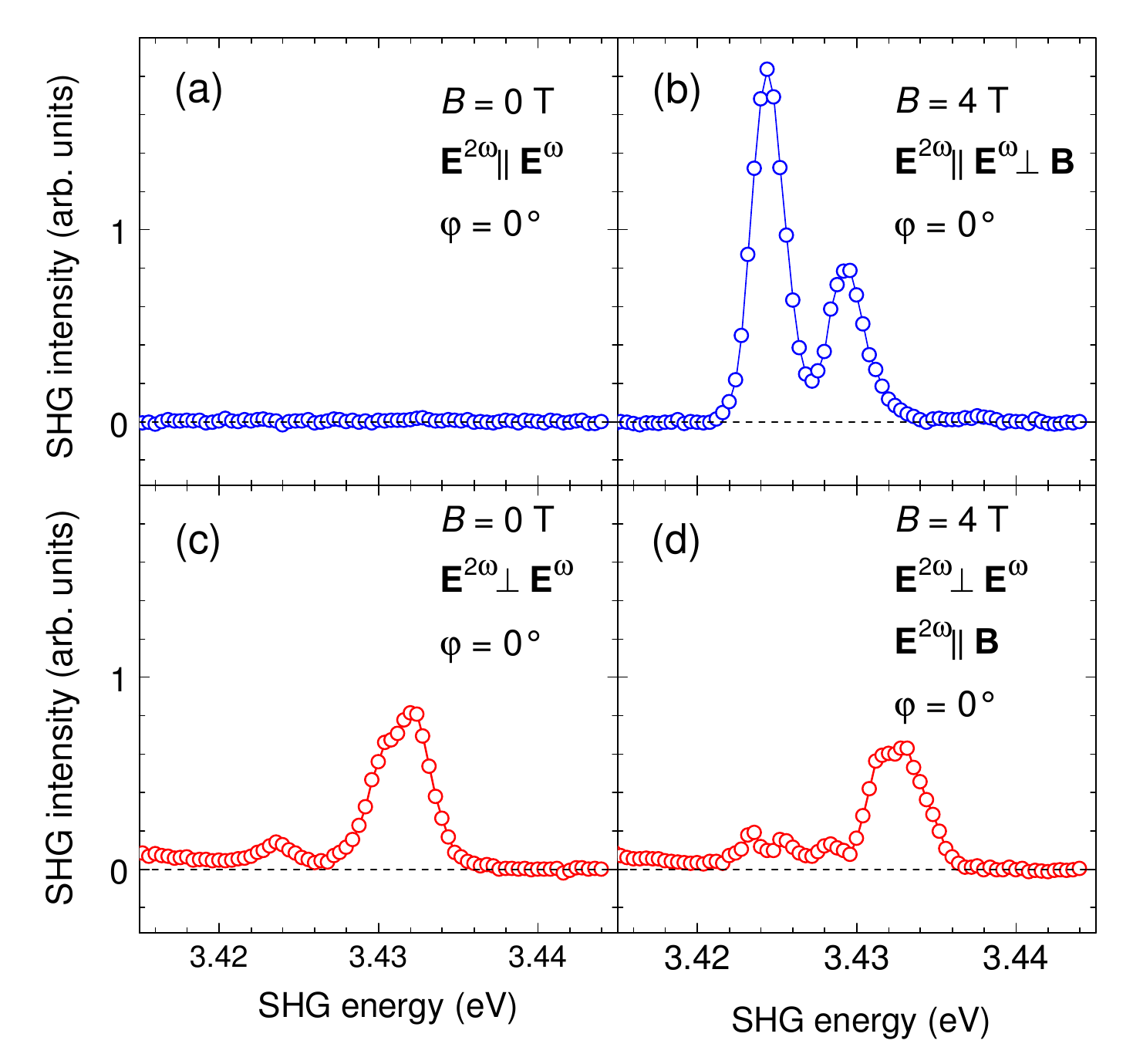}
\caption{(color online) Crystallographic and magnetic-field-induced
SHG in the spectral range of the $2s/2p(A,B)$ and $1s_L (C)$
excitons measured for the tilted geometry with
$\theta\approx45^\circ$. $T=1.6$~K. (a) There is no crystallographic
contribution to $I^{2\omega}_\parallel$ for
$\varphi(\mathbf{E}^\omega)=0^\circ$. (b) Pure magnetic-field-induced SHG
signals at $B=4$~T for $I^{2\omega}_\parallel$ and
$\varphi(\mathbf{E}^\omega)=0^\circ$ contributed by the $2s/2p(A,B)$
excitons. (c) and (d) $I^{2\omega}_\perp$ for the $1s_{\text{L}}(C)$
exciton does not change significantly from $B=0$ to $4$~T.}
\label{fig:figure12}
\end{figure}

It is instructive to compare the intensities of the crystallographic
and the magnetic-field-induced SHG signals. This comparison is
presented in Fig.~\ref{fig:figure12}, where all four panels,
recorded with the tilting angle $\theta\approx45^\circ$, have the
same intensity scale. In absence of magnetic field the strongest
crystallographic SHG signal is found for
$\mathbf{E}^{2\omega}\parallel\mathbf{E}^\omega$ and
$\varphi(\mathbf{E}^\omega)=90^\circ$, while it vanishes for
$\mathbf{E}^{2\omega}\parallel \mathbf{E}^\omega$ and
$\varphi(\mathbf{E}^\omega)=0^\circ$; see Fig.~\ref{fig:figure4}(c).
Fig.~\ref{fig:figure12}(a) demonstrates, that even for a tilted
sample no signal is observed for $\mathbf{E}^{2\omega}\parallel
\mathbf{E}^\omega$, $\varphi(\mathbf{E}^\omega)=0^\circ$, but in
magnetic field strong signals appear for this configuration, see
Figs.~\ref{fig:figure10}(a) and \ref{fig:figure10}(b). Thus, for a tilted sample
Fig.~\ref{fig:figure12}(b) shows the intensity of a pure MFISH
signal for the $2s/2p(A,B)$ states, which is even more intense than
the strongest crystallographic signal $I^{2\omega}_\perp$ in
Fig.~\ref{fig:figure12}(c) observed for the $1s_{\text{L}}(C)$ with
$\mathbf{E}^{2\omega}\perp \mathbf{E}^\omega$ and
$\varphi(\mathbf{E}^\omega)=0^\circ$, see Fig.~\ref{fig:figure4}(c).
Consequently, we can conclude that the susceptibilities of MFISH and
crystallographic SHG have comparable values. On the other hand, the
$1s_{\text{L}}(C)$ state is not strongly modified by the MFISH
contributions, compare amplitudes in Figs.~\ref{fig:figure12}(c) and
~\ref{fig:figure12}(d).

\subsection{Temperature dependence}

\begin{figure}
\includegraphics[width=0.45\textwidth,angle=0]{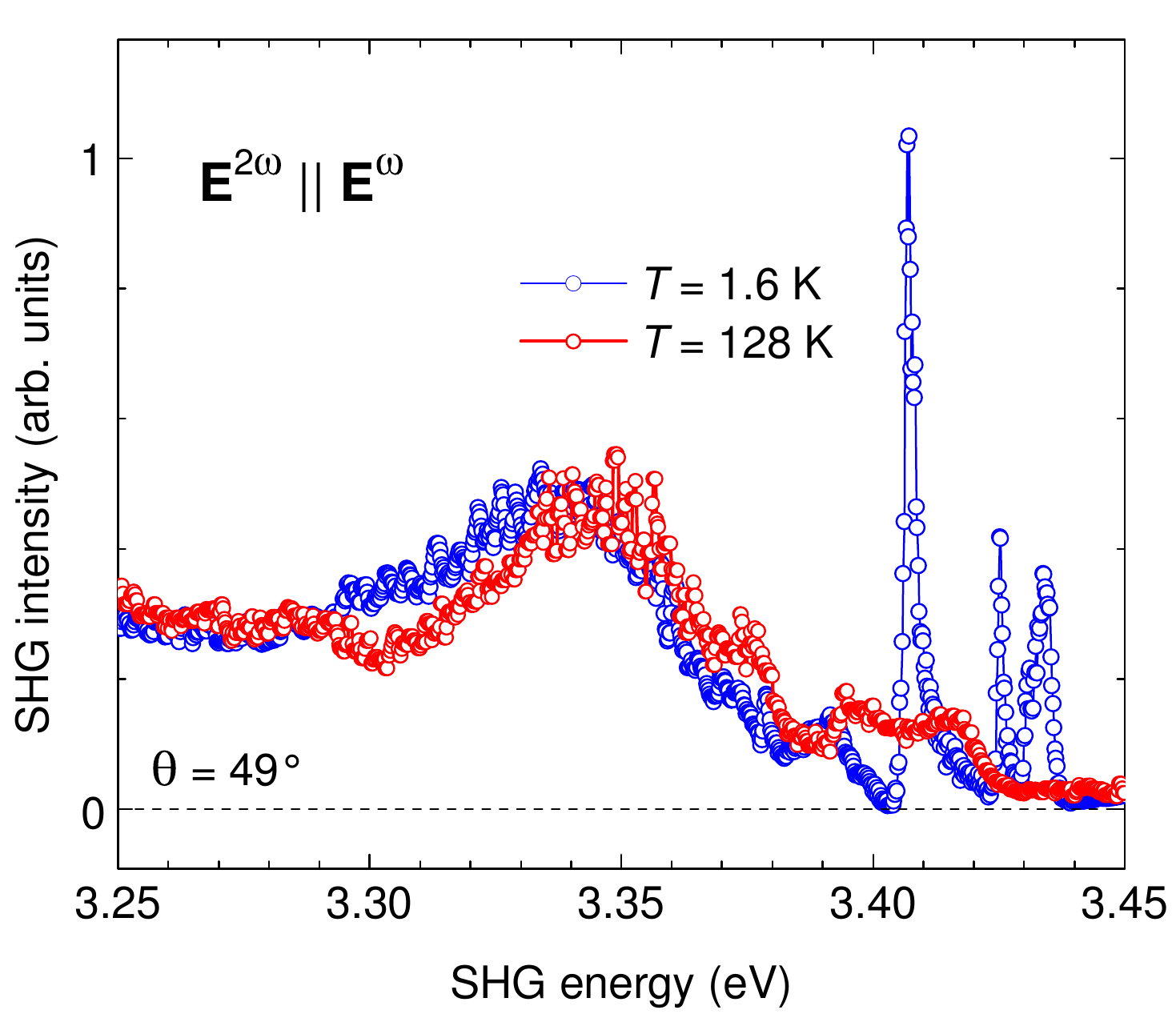}
\caption{(color online) Crystallographic SHG spectra of ZnO for
$\mathbf{E}^{2\omega}\parallel \mathbf{E}^\omega$ at $\varphi =
90^\circ$, measured at $T=1.6$~K (blue circles) and $128$~K (red
circles).}\label{fig:figure13}
\end{figure}

\begin{figure}
\includegraphics[width=0.45\textwidth,angle=0]{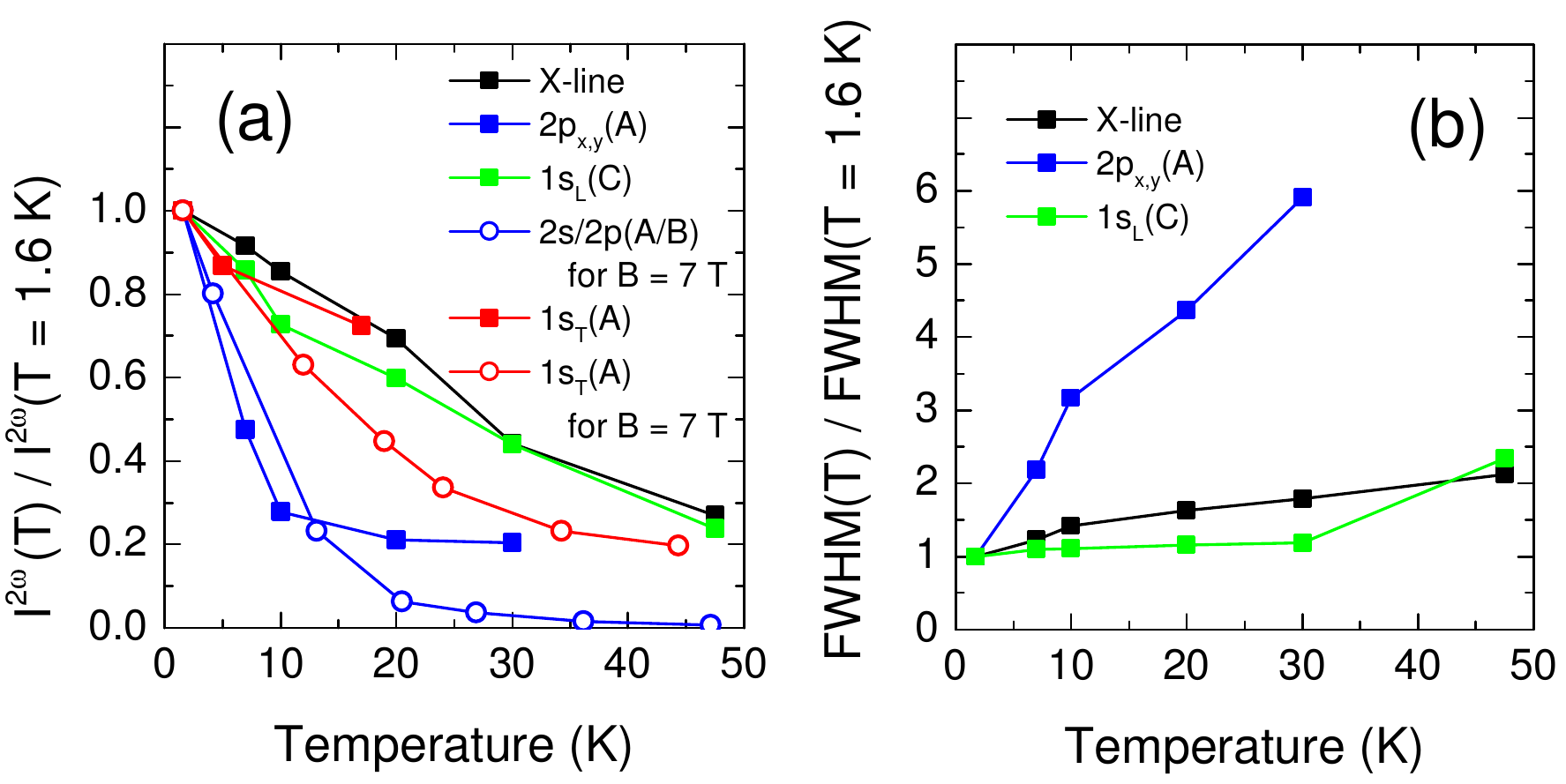}
\caption{(color online) (a) Normalized SHG intensity \emph{vs}.
temperature for different photon energies. Crystallographic signals
of the $1s_{\text{L}}(C)$ exciton (green squares) and the
unidentified $X$-line (black squares) decrease to about $20\%$ at
$T=50$~K. Crystallographic signal  of the $2p_{x,y}(A)$ state (blue squares) 
and magnetic-field-induced  signal of $2s/2p(A,B)$ states (open blue circles)
show a fast decay and vanish in the background for $T>30$~K. For
$B=7$~T the temperature dependencies of the $1s$ and $2s/2p$ states
change only slightly compared to the zero-field case, see open
symbols. (b) Normalized FWHM extracted from SHG measurements at
different energies. In contrast to the slow temperature increase for
the $1s_{\text{L}}(C)$ exciton (green squares) and the $X$-line
(black squares) the $2p_{x,y}(A)$ exciton (blue squares) shows a
rapid increase with temperature.} \label{fig:figure14}
\end{figure}

Figure~\ref{fig:figure13} compares the crystallographic SHG
intensities $I^{2\omega}_\parallel$ recorded at $1.6$~K and $128$~K.
While the off-resonant contribution has comparable intensity, the
exciton SHG signals strongly decrease with rising temperature. A
closer look at the detailed evolutions of the peak intensities shows
that all $1s$ state and the $X$-line intensities decrease slower
with temperature than the $2p_{x,y}(A)$ states, compare the results
shown in Fig.~\ref{fig:figure14}(a). At the same time, the full
width at half maximum (FWHM) of the $2p_{x,y}(A)$ line increases much
faster than those of the $1s_{\text{L}}(C)$ state and the $X$-line,
see Fig.~\ref{fig:figure14}(b). The magnetic field influences the
temperature dependence only slightly: the magnetic-field-induced signals show
a similar behavior as the crystallographic ones; the closed and open
dots in Figure~\ref{fig:figure14}(a) give the temperature
dependencies for zero field and $B=7$~T, respectively. We conclude
that the signal decays are rather independent of the SHG generating
mechanism. An explanation based on our theoretical model will be
discussed in Sec.~\ref{sec:discussion}.

\subsection{Joint action of magnetic and electric field}
\label{subsec:electric_SHG}

An applied electric field modifies the wave functions of the exciton
states and, therefore, offers another promising option for SHG
spectroscopy. The electric field is a polar vector of odd parity, in
contrast to the even parity magnetic field, so that it can mix
exciton wave functions of opposite parity.

\begin{figure}
\includegraphics[width=0.45\textwidth,angle=0]{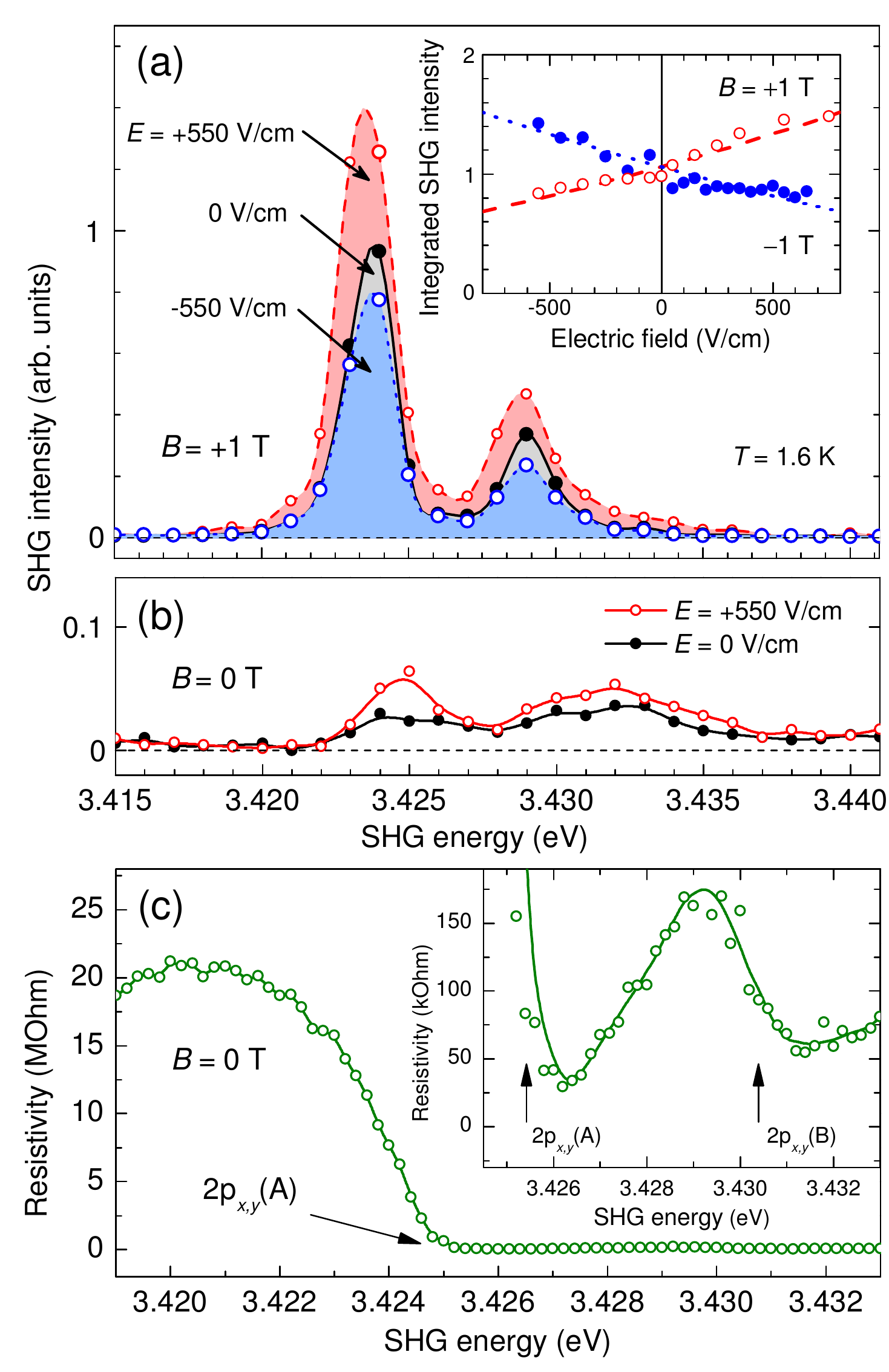}
\caption{(color online) (a) SHG spectra in the range of the
$2s/2p(A,B)$ excitons in ZnO subject to magnetic field and
combined electric and magnetic fields. An applied electric field of
$E=\pm550$~V/cm corresponds inside the crystal to the field
$E/\epsilon^\perp$, where $\epsilon^\perp=7.40$ is the static
dielectric permittivity perpendicular to the hexagonal $z$-axis.
Inset shows integrated intensity in the spectral range
$3.417-3.438$~eV as function of electric field for $B=+1$ and
$-1$~T. Symbols are experimental data and lines give best fits
according to $I^{2\omega}\propto(\pm B+\gamma E)^2$ with
$\gamma=2.5\times 10^{-4}$. (b) SHG spectra without applied magnetic
field. Black line shows residual crystallographic signal due to
slight sample misalignment. Red line demonstrates electric field
effect [intensity is increased by a factor of $4$ compared to (a)].
(c) Measured resistivity of the sample at $B=0$~T showing a strong
drop by about three orders of magnitude when $2\hbar\omega$ becomes
close to the $2p_{y,x}(A,B)$ excitons. Inset shows a close-up
($\times 100$) of this region. }\label{fig:figure15}
\end{figure}

The two spectra in Fig.~\ref{fig:figure15}(b) demonstrate the effect
of an applied electric field on SHG spectra for the $2s/2p(A,B)$
excitons in ZnO at zero magnetic field. In the absence of electric
field the residual SHG amplitude (about 1\% of those previously
discussed) originates from small strain caused by the sample holder
with electrical contacts. Application of an electric field
perpendicular to the $z$-axis $\mathbf{E} \perp \mathbf{k}
\parallel \mathbf{z}$ leads to an increase in the SHG signal. The electric
field effect is much more pronounced in combination with an applied
magnetic field $\mathbf{E} \perp \mathbf{B}$, see
Fig.~\ref{fig:figure15}(a). Here the SHG amplitude is initially
gained by applying a magnetic field of $+1$~T and then tuned by
adding an electric field of $\pm 550$~V/cm. The SHG signal increases
for positive electric fields and decreases for negative fields. The
rotational anisotropies of these signals are not changed by the
electric field. The variation of the integral SHG intensity with
electric field strength is shown in the inset of
Fig.~\ref{fig:figure15}(a). Such a behavior, that a weak effect
(here induced by the electric field) is enhanced when combined with
a stronger effect (by the magnetic field) is known in SHG
spectroscopy when these effects interfere with each other, see e.g.
Eq.~(3) in Ref.~\cite{Kaminski}.

Resistivity measurements have shown, that the incident laser beam
reduces the sample resistivity enormously (by several orders of
magnitude), when twice the fundamental photon energy $2\hbar\omega$
comes close to that of the $2p$ exciton states, see
Fig.~\ref{fig:figure15}(c). The resistivity is not instantaneously
restored when the laser is switched off.

\section{Theory of SHG at exciton resonances}
\label{sec:theory}
\subsection{General consideration}
\label{subsec:GC-theory}

Theoretical studies of SHG were performed for many model
semiconductors, see for example
\cite{Sipe1,Sipe2,Sipe3,Sipe4,Sipe5,Sipe6,Sipe7,Rashkeev1,Rashkeev2,Rashkeev3,Rashkeev4}.
These publications analyze the generation of the second and higher
harmonics by band theory or first-principle calculations, while
exciton contributions have remained essentially unexplored. The
complex experimentally observed exciton SHG signals for ZnO in
external fields as reported here demands development of a
corresponding microscopic theory.

In this section we focus on SHG effects in resonance with exciton
states. This requires the analysis of the wave function symmetries
for different exciton states and their modifications in external
magnetic and electric fields. We present a theoretical analysis for
the excitons in ZnO with wurtzite-type crystal structure. The
developed theoretical approach, however, can be readily applied to
other semiconductors. In particular, most of the suggested
mechanisms of magnetic- and electric-field-induced SHG at the
exciton resonances should prevail also in other materials.

To analyze SHG in close vicinity of an exciton resonance we write
the nonlinear optical susceptibilities $\chi_{ijl}({\cal
E}_{\mathrm{exc}},\mathbf{k}_{\mathrm{exc}},\mathbf{B},\mathbf{E})$
introduced in Eq.~(\ref{eq:P3}) for each exciton energy ${\cal
E}_{\mathrm{exc}}=2\hbar\omega$ in general form as
\begin{widetext}
\begin{eqnarray}
\chi_{ij\,l}({\cal
E}_{\mathrm{exc}},\mathbf{k}_{\mathrm{exc}},\mathbf{B},\mathbf{E})
\propto \sum_v \frac{\me{G|\hat
V^{2\omega}_i|\Psi_{\mathrm{exc}}}\me{\Psi_{\mathrm{exc}}|\hat
V^{\omega}_j|\psi_{v}}\me{\psi_{v}|\hat V^{\omega}_l|G}}{({\cal
E}_{\mathrm{exc}}-2\hbar \omega-i\Gamma_\mathrm{exc})({\cal
E}_v-\hbar
\omega)} \, \approx \frac{i} {\Gamma_\mathrm{exc}}
\me{G|\hat V^{2\omega}_i|\Psi_{\mathrm{exc}}}
M_{\mathrm{exc},G}^{2\text{ph}} \, . \label{chigen}
 \end{eqnarray}
 \end{widetext}
Here $\ket{G}$ denotes the unperturbed ground state with zero
energy, $\ket{\psi_v}$ describes intermediate virtual states,
$\ket{\Psi_{\mathrm{exc}}}$ describes the exciton state, and
$\Gamma_\mathrm{exc}$ is the exciton damping constant. The summation
in Eq.~(\ref{chigen}) is carried out over all intermediate states
satisfying the symmetry selection rules for the two-photon
transition from the ground to the exciton state described by the
matrix element $M_{\mathrm{exc},G}^{2ph}$.

To account for the effects of the external electric and magnetic
fields we consider the geometry where the crystallographic SHG
signals are suppressed, namely
$\mathbf{k}_{\mathrm{exc}}\parallel\mathbf{k}\parallel \textbf{z}$,
$\mathbf{B}\parallel \textbf{x}$, and $\mathbf{E}\parallel
\textbf{y}$. In this case the incoming field is described by
$\mathbf{E}^{\omega} = (E_x^\omega,E_y^\omega,0)$ so that the
outgoing polarization can be written as
$\mathbf{P}^{2\omega}_{\mathrm{eff}}=(P_{\mathrm{eff},x}^{2\omega},P_{\mathrm{eff},y}^{2
\omega},0)$.

The perturbation caused by the photon field $\mathbf{E}^\omega(\mathbf{r},t)$
is described by $(ie/m_0 \omega)[\mathbf{E}^\omega (\mathbf{r},t)  \mathbf{\hat p}]$,
where $e$ and $m_0$ are the charge and mass of free electron and $\mathbf{\hat p}$
is the momentum operator. Then the perturbation $\hat V_{x(y)}^{\omega}$  is given by
\begin{eqnarray}
\hat V_{x(y)}^{\omega} = \frac{ie}{m_0 \omega} {\hat p}_{x(y)} \exp(ik_z r_z) \approx \frac{ie}{m_0 \omega}
{\hat p}_{x(y)}(1+ik_z r_z+....)\, \label{Vxy},
\end{eqnarray}
where  ${\hat p}_{x(y)}$ are the projections of the momentum
operator $\mathbf{\hat p}$ on the light polarization components, $x$
or $y$, respectively. For $\hat V_{x(y)}^{2\omega}$ one should
substitute $k_z$ by $2k_z$ in Eq.~(\ref{Vxy}) and everywhere below,
as well as $\omega$ by $2\omega$. We are interested in the lowest
order effects in $k_z$, that is zero-order independent of $k_z$, if
it exists, or first-order, linear in $k_z$. Therefore, we keep only
two terms in the expansion of $\exp(ik_z r_z) = 1 + i k_z r_z$ and
consider the matrix elements of the form $\hat V_{x(y)}^{\omega} =
\hat D_{x(y)}^{\omega} + \hat Q_{x(y)z}^{\omega}$. The first term
$\hat D_{x(y)}^{\omega}=(ie/m_0\omega){\hat p}_{x(y)}$ corresponds
to the electric-dipole (ED) approximation for which the matrix
elements can be replaced \cite{Park} by the matrix elements of the
dipole operator $e{r}_{x(y)}/\hbar$. The operator $\hat
Q_{x(y)z}^{\omega} = -(ek_z/m_0\omega) {\hat p}_{x(y)}r_z$ includes
the electric-quadrupole (EQ) and the magnetic-dipole (MD)
contributions where the matrix elements can be replaced \cite{Park}
by the sum of matrix elements of the electric-quadrupole operator
$\hat Q_{x(y)z}^{\omega,q} = -(i e k_z/2) r_{x(y)} r_z$ and the
magnetic-dipole operator  $\hat Q_{x(y)z}^{\omega,m} = \pm (e \hbar
k_z/2m_0 \omega) {\hat L}_{y(x)}$. Here $\mathbf{\hat L}$ is the
orbital momentum operator. Depending on the perturbation $\hat
V^{\omega}$ and $\hat V^{2\omega}$ involved in the two-photon
absorption and one-photon emission, respectively, we denote the
resulting three-photon SHG process as $X^{2w}Y^{\omega}Z^{\omega}$,
where the $X,Y,Z$ are either $D$ (ED) or $Q$ (EQ+MD). We emphasize,
that the presence of EQ or MD transition for one of the steps
either excitation or emission, leads to a linear dependence of the
susceptibility on $k_z$.

In ZnO, the direct ED transitions between the valence and conduction
band states are allowed. The strongest one-photon process for
$\mathbf{k}\parallel \textbf{z}$ are the excitation of the
($\emph{s}\times\Gamma_5$) states or the emission from them. The
respective matrix elements can be written as $D^{\omega,a}$ or
$D^{2\omega,a}$, where the index $a$ denotes "allowed" transition
within the ED approximation according to the notation of Elliot
\cite{elliot}. In contrast, the one-photon ED "forbidden"
transitions to the $2p$ excitons in noncentrosymmetric wurtzite
semiconductors like ZnO may occur because the valence and conduction
band states do not have pure even or odd parities. These transitions
are much weaker compared to the $s$ exciton transitions and can be
described by the matrix elements $D^{\omega,f}$ or $D^{2\omega,f}$,
where the index $f$ denotes the "forbidden" character within the ED
approximation \cite{elliot}. In the used geometry such "forbidden"
transitions are possible only for the $2p_{x,y}$ states, and not for
the $2p_z$ state. Alternatively, the one-photon emission from all
three $2p_x$, $2p_y$, and $2p_z$ states may occur due to 
magnetic-dipole transitions described by the matrix element $Q^{2\omega,m}$.

The strongest two-photon process in ZnO is the excitation of the
$2p$ exciton states. It is ED allowed, exploiting intermediate
virtual states in the valence or conduction band. Such process
involves a transition between valence and conduction band states and
another transition between $s$ and $p$ envelopes in the same energy
band. The relevant two-photon matrix element is
$M_{2p,G}^{2\text{ph}}\propto D^{\omega, a}D^{\omega, f}$. On the
other hand, the direct two-photon absorption by the $s$ exciton
states in non-centrosymmetric semiconductors may occur within the ED
approximation via the intermediate virtual states in remote bands
\cite{vs}. In this case the two-photon matrix element is
$M_{s,G}^{2ph} \propto D^{\omega, a} D^{\omega, a}$. However, such
processes are much weaker than those for the $2p$ excitons
\cite{ig}. Alternatively, the $s$ states can be excited in the
two-photon process when the first transition is a MD transition (or
ED transition of "forbidden" character) to the $2p_{x,y}$ states and
the second one is a ED transition of "forbidden" character between
the $s$ and the $p$ envelopes. In this case the two-photon matrix
element is $M_{s,G}^{2ph} \propto D^{\omega, f} Q^{\omega,m}$ (or
$M_{s,G}^{2ph} \propto D^{\omega,f}D^{\omega, f}$).

Important information on the symmetry of the exciton states involved
in the SHG process is provided by the SHG rotational anisotropies.
According to Eq.~\eqref{eq:P3} the SHG intensity is given by
\begin{equation}
I^{2\omega}_{\perp \mathbf{B}} \propto |\chi_{yyy} \cos^2\varphi +\chi_{yxx} \sin^2\varphi |^2 \,, \label{eq:anieq1}
\end{equation}
if $\mathbf{E}^{2\omega} \bot \textbf{B}$; and by
\begin{equation}
I^{2\omega}_{\parallel \mathbf{B}} \propto |\chi_{xxy} \sin (2\varphi) |^2 \,,\label{eq:anieq2}
\end{equation}
for the $\mathbf{E}^{2\omega} \| \textbf{B}$ geometry. Here
$\varphi$ is the angle between $\mathbf{E}^\omega$ and the $y$-axis.
Note that $\textbf{B} \bot \textbf{y}$. For parallel polarization
$\mathbf{E}^{2\omega}\| \mathbf{E}^{\omega}$ one obtains
\begin{equation}
I^{2\omega}_\| \propto \cos^2\varphi |\chi_{yyy} \cos^2\varphi + (\chi_{yxx}+ 2\chi_{xxy}) \sin^2\varphi |^2 \, .\label{eq:anieq3}
\end{equation}

In a hexagonal 6\emph{mm} crystal the relation $\chi_{yyy}=
\chi_{yxx}+ 2\chi_{xxy}$ is fulfilled \cite{Birss} so that
\begin{equation}
I^{2\omega}_\| \propto |\chi_{yyy} \cos\varphi|^2 \,.\label{eq:anieq4}
\end{equation}
For perpendicular polarization $\mathbf{E}^{2\omega} \perp
\mathbf{E}^{\omega}$ one finds
\begin{equation}
I^{2\omega}_\perp \propto \sin^2\varphi |(\chi_{yyy}-2\chi_{xxy})\cos^2\varphi + \chi_{yxx} \sin^2\varphi |^2 \, ,\label{eq:anieq5}
\end{equation}
and if the relation $\chi_{yyy}= \chi_{yxx}+ 2\chi_{xxy}$ is
fulfilled, then
\begin{equation}
I^{2\omega}_\perp \propto |\chi_{yxx} \sin\varphi|^2\,.\label{eq:anieq6}
\end{equation}
Below we will show, that a magnetic field applied perpendicular to
the hexagonal $z$-axis may reduce the symmetry of an exciton state
and, consequently, it violates  the relation $\chi_{yyy}=
\chi_{yxx}+ 2\chi_{xxy}$.

In the following subsections we will proceed with the analysis of
different specific mechanisms of the field-induced mixing of exciton
states and derive the corresponding nonlinear optical
susceptibilities. The results of this analysis are summarized in
Table~\ref{tab:Mechanisms}. Relations between $\chi_{yyy}$,
$\chi_{yxx}$, and $\chi_{xxy}$ allow one to model the rotational
anisotropies for each particular mechanism. We note that the
admixture of exciton states in applied fields may lead to the
dependence of the wave function $\Psi_{\mathrm{exc}}$ and the
respective energy ${\cal E}_{\mathrm{exc}}$ on the exciton wave
vector $\mathbf{k}_{\text{exc}}$, as well as on $B_x$ and $E_y$.
These complex perturbations may lead to a nonlinear dependence of
the susceptibilities $\chi_{ijl}({\cal
E}_{\mathrm{exc}},k_{\mathrm{exc}},B_x,E_y)$ on $B_x$, $E_y$  and
$k_z$, via $k_{\mathrm{exc}}=2nk_z$. They act in addition to those
arising from the second term in the expansion of $\exp(i k_z z)$
according to Eq.~(5).

\subsection{SHG and exciton Stark effect ($\mathbf{E} \bot \mathbf{k}$)}
\label{subsec:theory_electric}

Let us first consider the SHG signals induced by an external
electric field $E_y$ which mixes the $2s$ and $2p_y$ exciton states
of opposite parity due to the Stark effect for the $A$ and $B$
excitons. However, it does not affect their spin states. Two
polariton branches can be formed for each of the mixed $2s/2p$
exciton states. The new energies ${\cal E}_{\mathrm{exc}}={\cal
E}_{2s_\mathrm{T}/2p_y}^\pm $ for the transversal lower polariton
branch (LPB) and ${\cal E}_{\mathrm{exc}}={\cal
E}_{2s_\mathrm{L}/2p_y}^\pm$ for the transverse upper polariton
branch (UPB) are given in the Appendix. The resulting wave functions
of the mixed states in Eq.~\eqref{eq:S3} are constructed from the
$2s$ and $2p_y$ components. In this process, all matrix elements for
excitation and emission in Eq.~(\ref{chigen}) become allowed in the
ED approximation $\hat V^{\omega(2\omega)}_{x(y)} = \hat
D_{x(y)}^{\omega(2\omega)}$. We denote  the corresponding SHG as
$D^{2\omega}D^{\omega}D^{\omega}$ as shown in the first row of
Table~\ref{tab:Mechanisms}. The corresponding SHG signals can be
observed only when the incoming light has a nonzero component
$E_y^\omega \ne 0$ responsible for excitation of the $2p_y$ state.
Therefore, for this process $\chi_{yxx}=0$ and the resulting
electric-field-induced susceptibilities
$\chi_{yyy}=2\chi_{xxy}=2\chi_{xyx}$ are proportional to the product
of the wave function admixture coefficients in Eqs.~\eqref{eq:S4}
and \eqref{eq:S5}. They can be written as
\begin{widetext}
\begin{eqnarray}
\chi_{yyy}({\cal E}_{2s/2p_{y}}^\pm,k_{\mathrm{exc}},0,E_y) \propto
C_{2s}(E_y)C_{2p_y}(E_y) = \frac{3eE_{y}a_\mathrm{B}({\cal
E}_{2p_y}-{\cal E}_{2s/2p_{y}}^\pm)} {(3eE_{y}a_\mathrm{B})^2+({\cal
E}_{2p_y}-{\cal E}_{2s/2p_{y}}^\pm)^2} \, , \label{chistark1}
\end{eqnarray}
\end{widetext}
where $a_\mathrm{B}$ is the exciton Bohr radius. One sees that these
electric-field-induced susceptibilities do not depend on the
absolute value of $k_z$, however, the direction of $\mathbf{k}$
parallel to the $z$-axis is important. If the electric field
perturbation energy is much smaller than the zero-field splitting of
the exciton states, $|eE_{y}a_\mathrm{B}| \ll |{\cal E}_{2p}-{\cal
E}_{2s_{\mathrm{T(L)}}}|$, then the susceptibilities depend linearly
on $E_y$. However, for larger fields a saturation is expected
because for $|eE_{y}a_\mathrm{B}| \gg |{\cal E}_{2p}-{\cal
E}_{2s_{\mathrm{T(L)}}}|$ the susceptibilities become independent of
$E_y$.

\subsection{SHG and magnetic field effects on excitons ($\mathbf{E} \bot \mathbf{k} \| \textbf{z}$)}
\label{subsec:theory_magnetic}

The effect of an applied magnetic field \textbf{B} on excitons shows
more facets than an electric field. We will discuss several
mechanisms acting when the magnetic field is applied along the
\emph{x}-axis, $\textbf{B}=(B_x,0,0)$:

1. \emph{The spin Zeeman effect}, which mixes the exciton spin
states through a perturbation $\propto (\sigma_xB_x)$, where
$\sigma_x$ is the corresponding Pauli matrix.

2. \emph{The orbital Zeeman effect}, which affects the $p$ states
having nonzero envelope orbital momentum $L=1$ and mixes the $2p_z$
and $2p_y$ states by a perturbation $\propto (L_xB_x)$.

3. \emph{The magneto-Stark effect}
\cite{Samoilovich,Thomas0,Thomas,Gross,Lafrentz}. This effect arises
from the oppositely directed Lorentz forces acting on electron and
hole in a magnetic field during the exciton center-of-mass motion.
The resulting perturbation of the exciton wave function is
equivalent to the effect of an effective electric field
$\mathbf{E}_{\mathrm{eff}}$ acting on the exciton at rest:
\begin{eqnarray}
\mathbf{E}_{\mathrm{eff}}=\frac{\hbar }{M_{\mathrm{exc}}}\left[
\mathbf{k_{\mathrm{exc}}}\times \mathbf{B} \right].
\label{Magneto-stark}
\end{eqnarray}
Here $M_{\mathrm{exc}}=m_{e}+m_{h}$ denotes the exciton
translational mass. In the given geometry the effective electron and
hole masses for motion parallel to the hexagonal $z$-axis
$\mathbf{k}_{\mathrm{exc}}\parallel \textbf{z}$ and
$\mathbf{E}_{\mathrm{eff}}
\parallel \textbf{y}$ have to be used: $m_{e}=m_{e}^{\parallel}$ and
$m_{h}=m_{h}^{\parallel}$.

The diamagnetic shift of the exciton energy occurs for all states
and is state-dependent \cite{Wheeler}. It does not directly lead to
a state mixing, but is can enhance mixing by other mechanisms due to
favorable energy shifts, bringing states closer to each other.

It is important, that the external magnetic field $B_x$ can mix
exciton states of different symmetry allowing two-photon resonant
excitation and one-photon resonant emission at a given energy and
thus leading to SHG signals. The Zeeman spin mixing may induce SHG
signals for one particular envelope exciton state. The orbital
Zeeman effect and the magneto-Stark effect mix states with different
envelope functions. The strength of this mixing depends on the
energy separation of these states at zero field. At a given exciton
energy the SHG signal might be induced by several mixing mechanisms
acting simultaneously. Below we analyze these mechanisms in detail
for each particular exciton state.

\subsubsection{Magnetic-field-induced SHG for $s$-type
excitons due to spin Zeeman effect} \label{subsubsec:Zeeman_s}

The spin states of the $s$-type excitons depend on the symmetries of
the conduction and valence bands. For excitons formed from the
conduction band of $\Gamma_7$ symmetry and the valence band of
$\Gamma_7$ symmetry  the resulting exciton states are of $\Gamma_5$,
$\Gamma_1$ and $\Gamma_2$ symmetry, split from each other by the
electron-hole exchange interaction. Examples of such states are the
$A$ and $C$ excitons in ZnO, or the $B$ and $C$ excitons in GaN. The
dipole-allowed $\Gamma_5$ state can occur for a one-photon process
and forms two polariton branches in the given geometry. The
$\Gamma_1$ state can become excited by a two-photon process if one
of the involved photons is due to the quadrupole perturbation or due
to the involvement of intermediate virtual states in remote bands
\cite{vs}. As a result, the $1s$ exciton states of the $A$ and $C$
excitons might be observed in the SHG spectrum due to the Zeeman
spin mixing of the $\Gamma_{5y}$ and $\Gamma_1$ states. The energies
of the new mixed polariton states ${\cal
E}_{{\Gamma_{5y}/\Gamma_1}}$ are given by Eq.~\eqref{eq:S8} in the
Appendix. The resulting wave functions of the mixed states described
by Eq.~\eqref{eq:S9} are constituted by both $\Gamma_{5y}$ and
$\Gamma_1$ components.

The two-photon excitation of the $s$ state might occur through an
electric-dipole/elecric-dipole ($D^{\omega,a}D^{\omega,a}$) or an
electric-dipole/magnetic-dipole ($D^{\omega} Q^{\omega,m}$) process
as discussed above. For the sake of clarity we consider the second
case in detail. It is represented by the process
$D^{2\omega}D^{\omega}Q^{\omega,m}$ in the third row of
Table~\ref{tab:Mechanisms}. Then the SHG process involves a matrix
element for the two-photon excitation with $\hat V^{\omega}_{x(y)} =
\hat Q^{\omega,m}_{x(y)z}$ for one of the photons and $\hat
V^{\omega}_{x(y)} = \hat D_{x(y)}^\omega$ for the second photon. The
subsequent one-photon ED emission with $\hat V^{2\omega}_y = \hat
D_y^{2\omega}$ is allowed through the $s_{\Gamma_{5y}}$ part of the
exciton wave function, so that $\chi_{xxy}=0$. The resulting
magnetic-field-induced nonzero susceptibilities
$\chi_{yyy}=\chi_{yxx}$ are given by
\begin{widetext}
\begin{eqnarray}
\chi_{yyy}({\cal
E}_{\Gamma_{5y}/\Gamma_1}^{\pm},k_\mathrm{exc},B_x,0) \propto
C_{\Gamma_5}(B_x)C_{\Gamma_1}(B_x) (k_z a_0)= \frac{2\mu_\mathrm{B}
g_{\mathrm{exc}}B_x({\cal E}_{\Gamma_5}-{\cal
E}_{\Gamma_{5y}/\Gamma_1}^{\pm})(k_z a_0)}{(\mu_\mathrm{B}
g_{\mathrm{exc}}B_x)^2+4({\cal E}_{\Gamma_5}-{\cal
E}_{\Gamma_{5y}/\Gamma_1}^{\pm})^2} \, .\label{chizeeman}
\end{eqnarray}
\end{widetext}
Here $g_\mathrm{exc}$ is the exciton $g$-factor, $a_0$ is the
lattice constant, and ${\cal E}_{\Gamma_5}^{\pm}$ is the zero-field
energy of the LPB or UPB exciton-polariton, respectively. The linear
dependence on $k_z$ enters through the matrix element of the
magnetic-dipole excitation with $\hat V^{\omega}_{x(y)} = \hat
Q^{\omega,m}_{x(y) z}$. If the exciton Zeeman splitting
$|\mu_\mathrm{B} g_{\mathrm{exc}}B_x|$ is much smaller than the
zero-field splitting of the corresponding exciton state, then the
susceptibilities depend linear on $B_x$ so that the SHG intensity
follows a $B^2$ dependence.

For $s$-symmetry  states, for which the SHG process is allowed by
the Zeeman spin effect, calculations show that
$\chi_{yyy}=\chi_{yxx} \ne 0$ and $\chi_{xxy} =\chi_{xyx}=0$. The
intensity of the SHG signal polarized perpendicular to the magnetic
field $I^{2\omega}_y \propto |\chi_{yyy}({\cal
E}_{\Gamma_{5y}/\Gamma_1},k_z^\omega,B_x,0)|^2$ does not depend on
the excitation polarization direction, while the signal polarized
parallel to the magnetic field vanishes: $I^{2\omega}_x \propto
|\chi_{xxy}({\cal E}_{\Gamma_{5y}/\Gamma_1},k_z,B_x,0)|^2 = 0$.
Simultaneously, for the parallel
$\mathbf{P}^{2\omega}_{\mathrm{eff}}
\parallel \mathbf{E}^{\omega}$ and the crossed
$\mathbf{P}^{2\omega}_{\mathrm{eff}} \bot \mathbf{E}^{\omega}$
geometries SHG signals of the same amplitude are predicted. Their
anisotropies are described by $I^{2\omega}_\parallel \propto
|\chi_{yyy}|^2 \cos^2\varphi$ and $I^{2\omega}_\bot \propto
|\chi_{yxx}|^2 \sin^2\varphi$.

For excitons formed by the conduction band of $\Gamma_7$ symmetry
and the valence band of $\Gamma_9$ symmetry the resulting exciton
states are of $\Gamma_5$ and $\Gamma_6$ symmetry. Examples are the
$B$ excitons in ZnO and the $A$ excitons in GaN. The Zeeman effect
mixes the dipole-allowed $\Gamma_5$ states and the dark $\Gamma_6$
state. In addition, one has to take into account the exchange
interaction between the $\Gamma_5$ components of the $A$ and $B$
excitons which may lead to SHG from the $\Gamma_{5y}/\Gamma_6$
exciton states with the same properties as described above for the
$\Gamma_{5y}/\Gamma_1$ excitons. In addition, the
$\Gamma_{5y}/\Gamma_6$ exciton states can be excited via the
$D^{\omega,a}D^{\omega,a}$ process.

\subsubsection{Magnetic-field-induced SHG for mixed $2s/2p$ excitons}
\label{subsubsec:magnetic_2s2p}

Let us now consider the effect of the effective electric field
$E_{\mathrm{eff}} = \frac{\hbar}{M_{\text{exc}}}k_\mathrm{exc} B_x$
on the exciton states. The mixing of $2s$ and $2p_y$ states of
opposite parity induced thereby is similar to that caused by the
Stark effect due to an external electric field. Simultaneously,
another type of mixing occurs due to the Zeeman orbital effect, but
this mechanism mixes the $2p_z$ and $2p_y$ states of the same
parity. The resulting energies ${\cal E}_{2s/2p_z/2p_y}^{i}$
($i=1,2,3$) of the mixed $2s/2p_z/2p_y$ polariton branches are
listed in Eq.~\eqref{eq:S13}. The appropriate wave functions in
Eq.~\eqref{eq:S14} are constructed from all three contributing
states with coefficients $C^i_{2s(2p_z,2p_y)}(B_x)$ given by
Eqs.~\eqref{eq:S15}-\eqref{eq:S17}. As a result, all mixed states
can be excited by two photons with polarization having a nonzero
field component $E_y^\omega \ne 0$ which excites the $\Psi_{2p_y}$
component. ED perturbations $\hat V^{\omega}_{x(y)} = \hat
D^{\omega}_{x(y)}$ are associated with the first photon and the
$\hat V^{\omega}_{y} = \hat D^{\omega}_{y}$ perturbation with the
second photon. There are three possible mechanisms allowing
observation of these mixed $2s/2p_z/2p_y$ states in one-photon
emission: (i) emission due to the $\Psi_{2s}$ component through the
ED perturbation $\hat V^{2\omega}_{x(y)} = \hat
D^{2\omega}_{x(y)}$; (ii) emission due to the $\Psi_{2p_z}$
component  through the magnetic-dipole perturbation $\hat
V^{2\omega}_{x(y)} = \hat Q^{2\omega,m}_{x(y)z}$; and (iii) emission
due to the $\Psi_{2p_y}$ component through the magnetic-dipole
perturbation $\hat V^{2\omega}_{x(y)} = \hat Q^{2\omega,m}_{x(y)z}$
or through the electric-dipole-forbidden process $\hat
V^{2\omega}_{y} = \hat D^{2\omega, f}_{yz}$.

\emph{a. Magneto-Stark effect}. In the first case (i) the mechanism
responsible for the SHG signal is coupling of the $2s$ and $2p_{y}$
states via the magneto-Stark effect and ED emission of the $2s$
state. The $D^{2\omega}D^{\omega}D^{\omega}$ process in the second
row of Table~\ref{tab:Mechanisms} corresponds to this mechanism. The
resulting magnetic-field-induced nonzero susceptibilities
$\chi_{yyy}=2 \chi_{xxy}=2\chi_{xyx}$ are proportional to the
product of admixed components in the corresponding wave functions:
\begin{eqnarray}
\chi_{yyy}^{2s/2p_{y}}({\cal E}^i_{2s/2p_z/2p_y},k_{exc},B_x,0)
\propto C^i_{2s}(B_x)C^i_{2p_y}(B_x).\, \label{chistark2}
\end{eqnarray}
The susceptibilities depend both on the magnetic field and the wave
vector value only via the dependence on the effective electric field
$E_{\mathrm{eff}} = \gamma k_\mathrm{exc} B_x$. This dependence is
linear when the energy of the effective electric field is smaller
than the zero field splitting of the states and it saturates in the
opposite limit.

\emph{b. Orbital Zeeman effect}. In the second case (ii) the
mechanism responsible for the SHG signal is due to coupling of the
$2p_z$ and $2p_{y}$ states via the orbital Zeeman effect and
magnetic-dipole emission from the $2p_z$ state. This is the
$Q^{2\omega,m}D^{\omega}D^{\omega}$ process shown in the fifth row
of Table~\ref{tab:Mechanisms}. This effect is expected to be much
weaker than the magneto-Stark effect and not important at the energy
where the $2s$ exciton is dominant. The resulting
magnetic-field-induced nonzero susceptibilities $\chi_{yyy}=2
\chi_{xxy}=2\chi_{xyx}$ can be written as
\begin{eqnarray}
\chi_{yyy}^{2p_{y}/2p_z}({\cal
E}^i_{2s/2p_z/2p_y},k_\mathrm{exc},B_x,0) \propto \nonumber \\k_z
a_0 C^i_{2p_z}(B_x)C^i_{2p_y}(B_x)\, . \label{chiyz}
 \end{eqnarray}
The linear dependence on $k_z$ comes from the matrix element of the
magnetic-dipole perturbation $\hat V^{2\omega}_{x(y)} = \hat
Q^{2\omega,m}_{x(y)z}$. Linear dependence of the $\chi$ on $B_x$ is
expected only for weak magnetic fields.

The rotational anisotropy patterns of the SHG intensities for the
processes induced by the magneto-Stark and the orbital Zeeman effect
are similar. The main feature for both of them is disappearance of
the signal in crossed geometry $\mathbf{P}_{\mathrm{eff}}^{2\omega}
\bot \mathbf{E}^\omega$ because $I^{2\omega}_\bot({\cal
E}^i_{2s/2p_z/2p_{y}}) \propto |\chi_{yxx}|^2=0$ for any
polarization direction of the excitation light $\mathbf{E}^\omega$.
The SHG signal in the parallel geometry
$\mathbf{P}_{\mathrm{eff}}^{2\omega}
\parallel \mathbf{E}^\omega$ can be modeled as $I^{2\omega}_\parallel
\propto |\chi_{yyy}|^2 \cos^2\varphi$, while the signal polarized
along the magnetic field direction varies as $I^{2\omega}_x \propto
|\chi_{xxy}|^2\sin^2(2\varphi)$. Since $\chi_{yxx}=0$, the signal
polarized perpendicular to the magnetic field direction can be
described by $I^{2\omega}_y \propto |\chi_{yyy}|^2\cos^4\varphi$.

\subsubsection{SHG due to spin Zeeman effect on $p$ states}
\label{subsubsec:spin_Zeeman}

\emph{a. The $p_y$ state}. For the $2p_y$ state the spin Zeeman
effect provides only one nonzero susceptibility due to the process
$Q^{2\omega,m}D^{\omega}D^{\omega}$ shown in the fourth row of
Table~\ref{tab:Mechanisms}. This susceptibility can be written as
\begin{eqnarray}
\chi_{yyy}^{2p_{y}}({\cal E}^i_{2s/2p_z/2p_y},k_\mathrm{exc},B_x,0)
\propto \frac{1}{2}k_za_0 C^i_{2p_y}(B_x),\,\label{chipy}
\end{eqnarray}
where the linear dependence on $k_z$ comes from the matrix element
of the magnetic-dipole perturbation $\hat V^{2\omega}_{x(y)} = \hat
Q^{2\omega,m}_{x(y)z}$.  We have neglected here the electron-hole
spin exchange splitting of the $p$ states. The magnetic field mixing
of the $2s/2p_z/2p_y$ envelopes reduces the SHG signal from the
Zeeman effect.

The corresponding rotational anisotropy patterns for the $2p_y$
states are: $I^{2\omega}_\|\propto |\chi_{yyy}|^2 \cos^6\varphi$,
$I^{2\omega}_\perp\propto |\chi_{yyy}|^2 \cos^4\varphi
\sin^2\varphi$, $I^{2\omega}_{\parallel\mathbf{B}}=0$ and
$I^{2\omega}_{\perp\mathbf{B}}\propto|\chi_{yyy}|^2\cos^4\varphi$.
Note, that the relation $\chi_{yyy}=\chi_{yxx}+2\chi_{xxy}$ valid
for 6\emph{mm} crystals \cite{Birss} is broken because of the
symmetry reduction by the magnetic field $B_x$. This field, directed
perpendicular to the $z$-axis, lifts the degeneracy of the $2p_x$
and $2p_y$ states.

\emph{b. The $p_x$ state}. A similar spin Zeeman mechanism is acting
on the $2p_x(A)$ exciton state, which is not mixed by the magnetic
field with other $p$ or $s$ states. The $p_x$ state can be excited
by two photons with a polarization with nonzero field component
$E_x^\omega \ne 0$ via the dipole perturbations $\hat
V^{\omega}_{x(y)} = \hat D^{\omega}_{x(y)}$ for the first photon and
$\hat V^{\omega}_{x}= \hat D^{\omega}_{x}$ for the second photon or
\emph{vice versa}. Emission is due to the Zeeman mixing of the
$\Gamma_5$ and $\Gamma_1$ spin states and the magnetic-dipole
perturbation $\hat V^{2\omega}_{x(y)} = \hat Q^{2\omega,m}_{x(y)z}$;
the relevant process $Q^{2\omega,m}D^{\omega}D^{\omega}$ is shown in
the fourth row of Table~\ref{tab:Mechanisms}. In fact it corresponds
to the very same Zeeman mixing of spin states as for the $1s$
states. Differences occur in the two-photon absorption and the
one-photon emission for the $s$ and $p$ states. The nonzero
susceptibility $\chi_{xxy}=\chi_{xyx}$ relevant to these processes
can be written as
\begin{eqnarray}
\chi_{xxy}({\cal E}_{2p_x}) \propto \frac{1}{2} k_za_0 .
\label{chipx}
\end{eqnarray}
where the linear dependence on $k_z$ comes from the matrix element
of the magnetic-dipole perturbation $\hat V^{2\omega}_{x(y)} = \hat Q^{2\omega,m}_{x(y)z}$.

There is no explicit dependence on the magnetic field in
Eq.~(\ref{chipx}). However, a finite $B_x$ is required for mixing
the $\Gamma_5$ and $\Gamma_1$ spin states because without magnetic
field the effect vanishes. $B_x$ does not appear in
Eq.~(\ref{chipx}) because we have neglected the exchange splitting
of the mixed states, so that they have the same energy. However,
even a very weak but finite magnetic field can mix them in equal
strength. In other words,  a kind of "phase transition" takes place,
as there is no effect at $B=0$ which then emerges instantaneously as
soon as $B>0$ with a finite magnitude independent of $B$. One may
compare this case with Eq.~(\ref{chizeeman}), where there is a
mixing of the $\Gamma_5$ and $\Gamma_1$ states that are split at
$B=0$. The magnetic field dependence in Eq.~(\ref{chipy}) comes only
from the $B$-dependence of the contribution of the $2p_y$ component,
given by $C_{2p_y}(B_x)$.

The SHG rotational anisotropy patterns for the $2p_x$ states via the
spin Zeeman mixing are: $I^{2\omega}_\|
\propto|\chi_{xxy}|^2\sin^2(2\varphi) \sin^2\varphi$,
$I^{2\omega}_\perp \propto|\chi_{xxy}|^2\sin^2(2\varphi)
\cos^2\varphi$,
$I^{2\omega}_{\parallel\mathbf{B}}=|\chi_{xxy}|^2\sin^2(2\varphi)$,
and $I^{2\omega}_{\perp\mathbf{B}}=0$. Note, that also here the
symmetry relation $\chi_{yyy}=\chi_{yxx}+2\chi_{xxy}$ valid for
6\emph{mm} hexagonal crystal \cite{Birss} is broken because of the
symmetry reduction in presence of a magnetic field $B_x$ directed
perpendicular to the $z$-axis. This field lifts the degeneracy of
the $2p_x$ and $2p_y$ states.

\begin{figure*}
\includegraphics[width=0.9\textwidth ,angle=0,keepaspectratio=true]{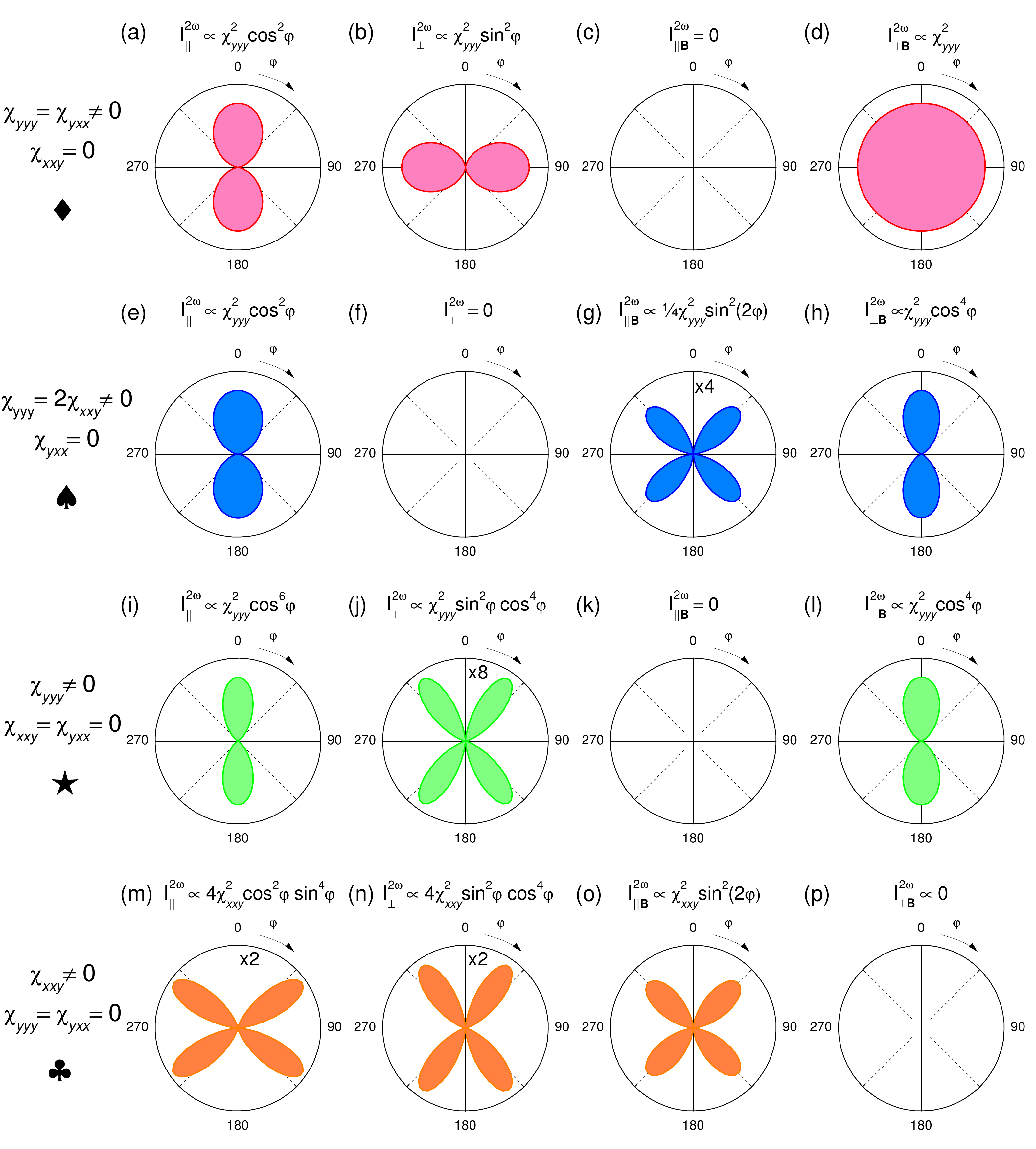}
\caption{(color online) Theoretical predictions for rotational
anisotropies of the SHG signal due to different contributions for
$I_\parallel\mapsto\mathbf{E}^{\omega}\parallel\mathbf{E}^{2\omega}$,
$I_\perp\mapsto\mathbf{E}^{\omega}\perp\mathbf{E}^{2\omega}$,
$I_{\parallel\mathbf{B}}\mapsto\mathbf{E}^{2\omega}\parallel\mathbf{B}$
and $I_{\perp\mathbf{B}}\mapsto\mathbf{E}^{2\omega}\perp\mathbf{B}$
according to Eqs.~\eqref{eq:anieq3}, \eqref{eq:anieq5},
\eqref{eq:anieq2}, and \eqref{eq:anieq1}, respectively, using the
relations from Table~\ref{tab:Mechanisms}: (a)-(d) spin Zeeman
effect $\blacklozenge$. (e)-(h) Stark effect / Magneto-Stark effect
/ orbital Zeeman effect $\spadesuit$. (i)-(l) spin Zeeman effect for
$2p_y$ $\bigstar$. (m)-(p) spin Zeeman effect for $2p_x$
$\clubsuit$.} \label{fig:figure16}
\end{figure*}

All mechanisms considered above for field-induced SHG at the $s$ and
$p$ exciton resonances are summarized in the
Table~\ref{tab:Mechanisms}. The susceptibilities are presented for
the geometry $\mathbf{k}
\parallel \textbf{z}$, $\mathbf{B}
\parallel \textbf{x}$, and $\mathbf{E} \parallel \textbf{y}$.
Resulting rotational anisotropy patterns corresponding to the
different mechanisms are illustrated in Fig.~\ref{fig:figure16}.

We have to note, without going into details, that the considered
mechanisms do not work in the Faraday geometry $\mathbf{B} \parallel
\mathbf{k} \parallel \textbf{z}$. A magnetic field applied along the
hexagonal $z$-axis does not mix the $\Gamma_1$ and $\Gamma_5$
states, therefore the mechanisms involving Zeeman spin mixing do not
induce SHG signals. The orbital Zeeman term $L_zB_z$ also does not
lead to admixture of the $2p_z$ states that is allowed for
quadrupole emission when $\mathbf{k}\| \textbf{z}$. The effective
electric field $E_\mathrm{eff}$ vanishes for the geometry
$\mathbf{B}\| \mathbf{k}$, so there is also no magneto-Stark effect.

\section{Discussion}
\label{sec:discussion}

\begin{table*}\footnotesize
\begin{tabular}{l|c|c|c|c|c}
Mechanisms & $1s$, $2s$  & $2s/2p_y$  & $2p_z/2p_y$  &   $2p_y$ & $2p_x$ \\
\hline \hline
Stark effect & & $\chi_{yyy}= 2\chi_{xxy}\ne 0$,& & & \\
$D_i^{2\omega}D_j^\omega D_l^\omega$& & & & & \\
$E_y \neq 0$, $B_x = 0$ & & $\chi_{yxx}=0$ $\spadesuit$ & & & \\
\hline
Magneto-Stark effect & & $\chi_{yyy}= 2\chi_{xxy}\ne 0$,& & & \\
$D_i^{2\omega}D_j^{\omega}D_l^{\omega}$& & & & & \\
$E_y = 0$, $B_x \neq 0$ & & $\chi_{yxx}=0$  $\spadesuit$ & & &\\
\hline
Spin Zeeman effect& $\chi_{yyy}= \chi_{yxx}\ne 0$, & & & &  \\
$D_i^{2\omega}D_j^{\omega}Q_l^{\omega,m}$& & & & & \\
$E_y = 0$, $B_x \neq 0$ & $\chi_{xxy}=0$ $\blacklozenge$ & & & & \\
\hline
Spin Zeeman effect & & & &   $\chi_{yyy} \ne 0$, $\bigstar$ & $\chi_{xxy} \ne 0$, $\clubsuit$ \\
$Q_i^{2\omega,m}D_j^{\omega}D_l^{\omega}$& & & & & \\
$E_y = 0$, $B_x \neq 0$ & & & &   $\chi_{xxy}= \chi_{yxx}=0$ & $\chi_{yyy}= \chi_{yxx}=0$\\
\hline
Orbital Zeeman effect & & & $\chi_{yyy}= 2\chi_{xxy}\ne 0$, & &\\
$Q_i^{2\omega,m}D_j^{\omega}D_l^{\omega}$& & & & & \\
$E_y = 0$, $B_x \neq 0$ & & & $\chi_{yxx}=0$ $\spadesuit$  & &
\end{tabular}
\caption{Different mechanisms providing SHG in external electric and
magnetic fields at the $1s(A,B,C)$, $2s(A,B,C)$ and $2p(A,B)$
exciton resonances in ZnO. Experimental geometry: $\mathbf{k}
\parallel z$, $\mathbf{E}=(0,E_y,0)$, and $\mathbf{B}=(B_x,0,0)$.
Due to symmetry $\chi_{xxy}=\chi_{xyx}$.} \label{tab:Mechanisms}
\end{table*}

The observed magnetic- and electric-field-induced SHG are specific
to excitons, no induced signals are observed in the off-resonant
energy range. In the following we discuss the various interactions
between light fields and exciton states, leading to the different
types of symmetry breaking presented in Sec.~\ref{sec:theory} in
order to explain our experimental data of Sec.~\ref{sec:results}. As
a general rule, we note that mixing of states with different
symmetries is the key to induce resonant nonlinear susceptibilities.
Rotational anisotropy measurements of the SHG signals give
comprehensive information on the symmetry of the involved nonlinear
susceptibilities and, therefore, of the underlying origins. Thus, they
help to distinguish different nonlinear optical mechanisms, which is
especially important when more than one mechanism become
involved. Therefore, the measured rotational anisotropy patterns of
the SHG intensity should reflect the anisotropies predicted by our
theoretical considerations. For the sake of convenience, all
discussed states will be referred to as excitons, although states
that contain admixtures of $s$ envelope states can couple so
strongly to the light field that exciton-polaritons may be formed.

The energy shifts of SHG lines in magnetic field and the field
dependence of their intensities should be in accord with the theory
as well. The computed energies depend strongly on the input
parameters, which are the zero-field exciton energies and the
electron and hole effective masses. Here we use the zero-field
exciton energies from Refs.~\cite{Fiebig2,Fiebig3}: ${\cal
E}_{2s_T}=3.4227$~eV,  ${\cal E}_{2s_L}=3.4232$~eV,  ${\cal
E}_{2p_z}=3.4240$~eV, and ${\cal E}_{2p_{x,y}}=3.4254$~eV for the
$A$ series as well as ${\cal E}_{2s_T}=3.4276$~eV,  ${\cal
E}_{2s_L}=3.4304$~eV, ${\cal E}_{2p_z}=3.4292$~eV, and ${\cal
E}_{2p_{x,y}}=3.4303$~eV for the $B$ series of excitons. We use the
electron effective mass $m_e=0.27m_0$ and the following hole
effective masses obtained from first principles calculations
\cite{Lambrecht}: $m_\parallel = 2.74 m_0$, $m_\perp = 0.54 m_0$ for
the $A$ excitons and $m_\parallel = 3.03 m_0$, $m_\perp = 0.55 m_0$
for the $B$ excitons. The static dielectric constants
$\epsilon_\parallel=8.49$ and $\epsilon_\perp=7.40$ are taken from
Ref.~\cite{Yoshikawa97}. Spin Zeeman splitting is not included in
the calculations for the $2s/2p_z/2p_y$ states. The discussed
interactions act on states with the same spin wave functions and
thus the spin Zeeman splitting leads to an energy shift, that is the
same for all involved states. Therefore, even though the number of
lines in Fig.~\ref{fig:figure9} will increase, it should not lead to
additional contributions.

\subsection{SHG mechanism for $1s$ exciton}

The SHG signals at the $1s$ exciton resonance can be fully explained
by the spin Zeeman effect, mixing one-photon allowed orthoexcitons
and two-photon allowed paraexcitons leading to the relations
$\chi_{yyy}=\chi_{yxx}\neq0$ and $\chi_{xxy}=0$, see
Table~\ref{tab:Mechanisms} and Sec.~V.C.1. The nonzero component
$\chi_{yyy}$ given by Eq.~\eqref{chizeeman} predicts the rotational
anisotropies shown in Figs.~\ref{fig:figure16}(a)-(d) as well as a
quadratic dependence of the SHG intensity on the magnetic field
$I^{2\omega}\propto B^2$. The measured intensities
$I^{2\omega}_\parallel$ and $I^{2\omega}_\perp$ show indeed the
predicted $\cos^2\varphi$ and $\sin^2\varphi$ shape and also have
the same amplitude, see Fig.~\ref{fig:figure6}(c). Further, no
signal was detected for $I^{2\omega}_{\parallel \mathbf{B}}$.
$I^{2\omega}_{\perp \mathbf{B}}$ was found to be isotropic (not
shown). Figure~\ref{fig:figure7}(a) proves the square dependence of
$I^{2\omega}$ on the magnetic field strength. For the $1s(B)$
exciton, which emerges from a different valence band, two SHG
mechanisms may be active: (i) mixing of the $\Gamma_5$ and
$\Gamma_6$ states due to the spin Zeeman effect, and (ii) mixing of
the $A$ and $B$ valence bands due to exchange interaction.

\subsection{SHG mechanisms for $2s/2p$ excitons}

A similar process based on the magnetic-field-induced spin Zeeman
effect, mixing the $\Gamma_5$ and $\Gamma_1$ states of the $1s(A)$
exciton, is valid also for the $2s(A)$ excitons. However, the
comparison of Fig.~\ref{fig:figure6}(c) and
Figs.~\ref{fig:figure10}(a,b) shows that the observed anisotropies
of the $n=2$ signals are different from those of the $1s$ states:
For the $n=2$ $(A)$ excitons the shape of $I^{2\omega}_\perp$ cannot
be described by the form $\sin^2\varphi$, and
$I^{2\omega}_\parallel$ and $I^{2\omega}_\perp$ do not have the same
amplitude. These anisotropies need to be explained by other
mechanisms than the spin Zeeman effect on the $s$ envelope. In
Ref.~\cite{Lafrentz} we have shown, that the magneto-Stark effect is
the dominant mechanism in the $2s/2p(A,B)$ exciton region. It can
explain the shapes shown in Figs.~\ref{fig:figure10}(a),
\ref{fig:figure10}(c), and \ref{fig:figure10}(d), which closely
resemble the predicted shapes in Figs.~\ref{fig:figure16}(e),
\ref{fig:figure16}(g), and \ref{fig:figure16}(h), respectively. But
a closer look at the non vanishing $I^{2\omega}_\perp$ intensity
reveals, that the spin Zeeman effect on both the $s$ and $p$
envelopes add a small contribution, note the multiplication factor
$20$ in Fig.~\ref{fig:figure10}(b). The complex shape of
$I^{2\omega}_\perp$ in Fig.~\ref{fig:figure10}(b) can only be
explained by a combination of the spin Zeeman effects on the $n=2$
states. Their predicted shapes are shown in
Figs.~\ref{fig:figure16}(b) and \ref{fig:figure16}(j), respectively.
To compare their impact, Fig.~\ref{fig:figure10} and
Fig.~\ref{fig:figure11} give their ratio $a/b$. $a$ and $b$ describe
the relative strength of the spin Zeeman effect on the $s$ and $p_y$
envelope, respectively. Both contributions to $I^{2\omega}_\perp$
can be estimated to be of about 5\% of the peak intensity of
$I^{2\omega}_\parallel$, while it follows from the fit in
Fig.~\ref{fig:figure10}(b), that the spin Zeeman contribution from
the $2p_y(A)$ part is twice as strong as from the $2s(A)$ part. As a
consequence, the SHG induced by the magneto-Stark effect involving
only ED excitation as well as ED emission processes is about $~40$
times more efficient than the SHG due to the spin Zeeman mixing
utilizing a MD transition. The orbital Zeeman effect cannot be
distinguished from the magneto-Stark effect via anisotropies (see
non zero components of both contributions in
Table~\ref{tab:Mechanisms}) and might be of importance too, due to
the $2p_y/2p_z$ mixing, especially for the regions, where the $2p_z$
part is large. However, the probability of MD emission from the
$2p_z$ part is very low in comparison with the probability of ED
emission from the $2s$ state. Thus, we assume that the orbital
Zeeman effect does not play a leading role and the magneto-Stark
effect brings the dominant contribution to the SHG signal at the
$2s/2p$ exciton resonances.

Further, the weak nonzero signals in the crossed geometry that do
not follow from the magneto-Stark effect were also observed in the
range of the $n=2$ $B$ excitons. The anisotropy shapes for the $B$
series presented in Figs.~\ref{fig:figure11}(a),
\ref{fig:figure11}(b), \ref{fig:figure11}(e), and
\ref{fig:figure11}(f) show many similarities to those of the $A$
series: the strongest signals are observed for
$I^{2\omega}_\parallel$ with a twofold $\cos^2\varphi$ symmetry
pattern, whereas $I^{2\omega}_\perp$ is a mixture of a fourfold
$\sin^2\varphi\cos^4\varphi$ and a twofold $\sin^2\varphi$ pattern,
originating from the spin Zeeman effects. In addition, the
contribution $I^{2\omega}_\perp\propto \sin^2\varphi$ in the range
of the $n=2$ $B$ exciton energies might also arise from the spin
Zeeman effect on the $3s(A)$ or $1s(C)$ excitons, located in the
same spectral range. However, the $3s(A)$ contribution is rather
unlikely, because of the high main quantum number considerably
reduces its oscillator strength in comparison to $1s$ and $2s$
states. By contrast, the $1s(C)$ is also seen as strong, broad
feature in the crystallographic SHG spectra, see
Fig.~\ref{fig:figure3}. Actually an additional contribution from the
$1s(C)$ exciton would explain, why we observed a strong influence of
the `$s$' type spin Zeeman effect on the shape of the anisotropies
for $2\hbar\omega=3.432$~eV; see Figs.~\ref{fig:figure11}(e) and
(f). Only in this region the ratio $a/b$ is larger than $1$.
Further, the magneto-Stark effect is not as important as for the
other energies; compare the ratios
$I^{2\omega}_\parallel/I^{2\omega}_\perp$ at the three energies
shown in Fig.~\ref{fig:figure11}.

The strongest SHG signals are observed for $I^{2\omega}_\parallel$
and their spectral maxima follow the energies of the states with
dominant $2p_y(A,B)$ wave functions. In higher fields ($>6$~T) the
peaks follow lines with dominant $2s$ wave functions, because the
admixture of the $2p_z$ to $2p_y$ states reduces the efficiency of
the magneto-Stark induced SHG; see the black circles in
Fig.~\ref{fig:figure9}. The analysis of the anisotropies in the
regions where the $2p_y/2p_z$ wave functions dominate gives the same
results as the analysis of states with large $2s$ contributions, but
with less influence from the $s$-type spin Zeeman effect.
Consequently, the SHG at the $2s/2p_y/2p_z$ states is governed
mainly by the magneto-Stark effect.

The contributions from the spin Zeeman effect on the $2p_x$ states
are best seen in Fig.~\ref{fig:figure8} for $I^{2\omega}_{\parallel
\mathbf{B}}$ in the geometry $\varphi(\mathbf E^\omega)=45^\circ$
and $\mathbf{E}^{2\omega}\parallel \mathbf{B}$. The spectral maxima
follow the energies of the calculated energies of the $2p_x(A,B)$
exciton states; compare $2p_x$ lines and red circles in
Fig.~\ref{fig:figure9}. In addition, the ratio of the spectrally
integrated intensities $I^{2\omega}_{\parallel
\mathbf{B}}/I_\parallel^{2\omega}$ should be $1/4$ or less without
the contribution of the spin Zeeman effect for the $2p_x$ states. We
observed, however, a deviating ratio of about $1/3$ [compare area
under curves in Fig.~\ref{fig:figure8}(b)], underlining the
importance of the spin Zeeman mechanism for the $2p_x$ orbitals
oriented parallel to the magnetic field. Nevertheless, the
contributions of this mechanism to the $I_\parallel^{2\omega}$
signals are not significant and most of the integrated SHG intensity
in the $I^{2\omega}_{\parallel \mathbf{B}}$ spectra comes from the
$2s/2p_y/2p_z$ mixed states due to the magneto-Stark effect.

The developed model describes well the measured angular dependencies
of the SHG intensities and it is in reasonable accordance with the
energy shifts of the exciton states shown in Fig.~\ref{fig:figure9},
ensuring the validity of the presented SHG mechanisms.

Another intriguing feature of the discovered mechanisms is the
complex behavior of the spectrally integrated SHG intensity shown in
Fig.~\ref{fig:figure7}(b). The SHG signals from the magneto-Stark
and orbital Zeeman effects are expected to saturate when the related
energies become larger than the zero field splitting of the involved
exciton states. The typical values of the $|2s_\mathrm{T}-2p_y|$
exciton splitting is about $3$~meV for the $A$ and $B$ excitons in
ZnO. While the longitudinal-transverse splitting of the $2s(A)$
exciton is about $0.5$~meV, the respective splitting for the $B$
exciton is about $3$~meV. Thus, the saturation condition is reached
for the orbital Zeeman effect ($g_{\text{orb}}\mu_{\text{B}}B_x$)
around $B\approx8$~T, but it is not fulfilled for the magneto-Stark
effect ($3eE_{\text{eff}}(B)a_{\text{B}}$) even at the strongest
field of $B\leq10$~T. The spin Zeeman effect for the $2p_x$ state is
independent of magnetic field, whereas the susceptibility decreases
with the fraction of $C_{2p_y}$, as for the $2p_y$ state, compare
Eq.~\eqref{chipy}. Due to the linewidths of the exciton resonances,
we were not able to resolve in Fig.~\ref{fig:figure8} individual
lines, but rather the interplay of contributions from different
energies leading to the observed complex behavior of the spectrally
integrated intensity. For the geometry shown in
Fig.~\ref{fig:figure7}(b) we take into account the magneto-Stark and
orbital Zeeman effects to model the SHG intensity dependence. The
model calculation for the strongest peak $2\hbar\omega=3.4254$~eV
reproduces well the observed dependence, assuming $\Gamma=1.2$~meV
and $\chi^{\text{magneto-Stark}}:\chi^{\text{orbital
Zeeman}}\approx100:1$ . Slight deviations can be expected, as in the
experiment the data was spectrally integrated. If the spin Zeeman
effect is taken into account the model calculations lead to a
dependence with a shoulder, shown for $\mathbf{E}^{2\omega}\parallel
\mathbf{E}^\omega\perp \mathbf{B}$ in Ref.~\cite{Lafrentz}.

\subsection{Temperature dependence}

The temperature dependence of the integrated SHG intensity can be
qualitatively understood by a simple consideration.
Eq.~\eqref{chigen} shows that the susceptibility for resonant SHG
depends inversely linear on the exciton damping 
$\Gamma_\text{exc}$, which is contributed by inhomogeneous and homogeneous broadening of the exciton. The inhomogeneous broadening in the studied sample does not exceed 1 meV, one can see that in Fig.~\ref{fig:figure6}(a), where it is already limited by the laser spectral width. Therefore, at temperatures exceeding $10-20$~K the exciton linewidth is controlled by the homogeneous broadening due to scattering on acoustic phonons. This broadening increases linearly with the temperature. 
The SHG peak intensity, proportional to the
susceptibility, depends inversely quadratically on the
linewidth $I^{2\omega}\propto\Gamma_{\text{exc}}^{-2}$. The
dependencies for the exciton resonances in Fig.~\ref{fig:figure14}
are spectrally integrated, for which the intensities in principle
have to be multiplied with the linewidth leading to the integrated
SHG being proportional to $\Gamma_{\text{exc}}^{-1}$ and respectively to $T^{-1}$. In contrast,
the crystallographic SHG signal measured in the off-resonant region
remains constant with increasing temperature, compare blue and red curves in the lower energy region
in Fig.~\ref{fig:figure13}.

\subsection{Origin of $X$-line}

To our knowledge the observed strong SHG line at $3.407$~eV has not
been reported in linear absorption spectra studies. However, the
observed temperature and angular dependencies of the SHG at this
line correspond to those of the free $1s(C)$ exciton; compare
Figs.~\ref{fig:figure4}(b) and \ref{fig:figure4}(d). Thus, a
correlation between these resonances is likely. In addition, we want
to point out the recent observation of an unknown line at
$\sim3.405$~eV with polarization pattern and uniaxial pressure
coefficients matching those of the $C_T(\Gamma_1)$ state
\cite{Anna}, suggesting to link the $X$-line to the $C$ valence
band. We did not find resonances at the corresponding energies for
the $A$ and $B$ excitons. Clarification of the $X$-line origin needs
further investigations. We also note, that phase synchronization for
the fundamental and second harmonic waves are of great importance
for SHG. Therefore, the strong SHG line at $3.407$~eV might be due
to phase matching \cite{Haueisen} as consequence of polaritonic
effects in the dielectric function in the exciton energy range
\cite{Cobet} of ZnO.

\subsection{Crossed electric and magnetic fields}

To verify the magneto-Stark effect as dominant source of the
observed SHG signals, it is instructive to discuss the joint action
of external magnetic and electric fields. Theory predicts that an
electric field produces the same type of symmetry breaking as the
effective electric field induced by a magnetic field. This is proven
by the fact, that the anisotropy of magnetic-field-induced signals
is not changed by additionally applying an electric field.
Nevertheless, the electric field acts very differently in comparison
to the magnetic field. The effective electric field acts on exciton
levels only, whereas the electric field creates a potential
throughout the crystal.

Figure~\ref{fig:figure15}(b) shows, that an electric field of
$550$~V/cm has only a weak effect on the SHG intensity. The
interference of its action with the effective electric field induced
by a magnetic field of $1$~T in Figure~\ref{fig:figure15}(a) shows,
that the electric field gives indeed surprisingly small
contributions. The magnitude of the effective electric field is
$E_{\mathrm{eff}}=\frac{\hbar}{M_{\text{exc}}}
k_{\mathrm{exc}}B_{x}$. The theoretical value for the exciton
translational mass in ZnO is about $3m_{0}$ \cite{Lambrecht} and
$k_{\mathrm{exc}}=n\mathcal{E}_{\mathrm{exc}}/\hbar c\approx
0.03$~nm$^{-1}$ for $\mathcal{E}_{\mathrm{exc}}=3.425$ eV with a
refractive index $n\approx 1.97$ \cite{Bond65,Yoshikawa97}. Thus,
$E_{\mathrm{eff}}$ can be estimated as $\approx 12$~V/cm for
$B=1$~T. Consequently, the ratio between the effective electric
field and the electric field strength giving the same effect is
$\gamma=\frac{1}{12\text{[V/cm]}\epsilon _{\perp }}\approx
1.13\times 10^{-2}$~T/V (where $\epsilon _{\perp }$ is the relative
dielectric permittivity) and would have to be used for the fit
function $I^{2\omega }\propto (\pm B\pm \gamma E)^{2}$. Instead the
best fit to the data shown in the inset of Fig.~\ref{fig:figure15}
(a) was achieved for a value that is $50$ times smaller. Such a discrepancy evidences that much weaker electric field is in fact acting on the excitons in our experiments. Indeed, as it is shown in Fig.~\ref{fig:figure15}(c),  the sample resistivity was reduced drastically when the $2\hbar \omega$ of the laser light approaches $2p$ states  and after illumination the resistivity
is only slowly restored. We suggest that the screening of the
external field by carriers trapped in deep centers is responsible
for the observed discrepancy.

\section{Conclusions}
\label{sec:conclusions}

In summary, new exciton phenomena in bulk hexagonal ZnO have been
thoroughly studied by optical second harmonic generation in the
spectral range of the $1s(A,B)$, $2s(A,B)$, $2p(A,B)$, and $1s(C)$
excitons, both experimentally and theoretically. While symmetry
considerations forbid any crystallographic SHG for
$\mathbf{k}\parallel \mathbf{z}$, strong magnetic-field-induced
contributions are found for this geometry. Novel microscopic
mechanisms for these nonlinearities are identified and confirmed by
detailed experimental studies, addressing the magnetic field,
electric field, temperature and polarization dependencies of the SHG
signals. 

We present an in-depth theoretical analysis on the basis of
phenomenological and microscopic approaches, which suggests several
mechanisms induced by external magnetic field. The magnetic field
produces a multifaceted action, depending on the exciton type. The
nonlinear mechanisms are related to the spin and orbital Zeeman
effects, and to the magneto-Stark effect. For $1s(A,B)$ excitons the
main mechanism of magnetic-field-induced SHG is related to the spin
Zeeman effect, which mixes the different spin wave functions. On the
other hand, the mixing of envelope wave functions of opposite parity
by the magneto-Stark effect due to an effective electric field is
the key mechanism for magnetic-field-induced SHG at the closely
spaced $2s/2p(A,B)$ excitons. The role of the orbital Zeeman effect
for mixing of the $2p_z$ and $2p_y$ orbitals and the spin Zeeman
effect on the $2p_x$ and $2p_y$ spin wave functions has been also
discussed. Application of an external electric field gives rise to
the Stark effect enabling SHG by mixing the wave functions of the
$2s/2p(A,B)$ excitons. 

We show the key importance of magnetic-
and electric-field-induced symmetry reductions for inducing
nonlinearities in bulk hexagonal ZnO, a phenomenon which should
occur in the same way also for other material systems. Tailoring
these symmetry reductions of the exciton level structure opens new
degrees of freedom in the nonlinear spectroscopy of excitons.

\begin{acknowledgments}
The authors are thankful to D.~Fr\"ohlich for highly stimulating
discussions and help with experiments, E.~L.~Ivchenko for
discussions, and to B.~Kaminski for his contribution to initial
stage of this study. This work was supported by the Deutsche
Forschungsgemeinschaft and the Russian Foundation for Basic
Research. V. V. P. research stay in Dortmund was supported by the
Alexander-von-Humboldt Foundation.
\end{acknowledgments}

\appendix
\section*{Appendix: Theoretical consideration of exciton state mixing in electric and magnetic fields}
\label{sec:Appendix}

Here we describe the mixed exciton states in hexagonal ZnO subject
to external electric or magnetic fields. Only the geometry used in
the experimental part of this paper is analyzed:
$\mathbf{k}\parallel \textbf{z}$, $\mathbf{E}=(0,E_y,0)$, and
$\mathbf{B}=(B_x,0,0)$.

\subsection{Exciton states in electric field perpendicular to hexagonal $z$-axis.}

The external electric field $E_y$ mixes the $2s$ and $2p_y$ exciton
states of opposite parity for the $A$ and $B$ exciton series, but
does not affect their spin states. We neglect for simplicity the
interaction between the $A$ and $B$ series and consider them
independently within the polariton concept. Then for each series the
exciton eigenenergies in the external field $E_y$ and the mixed
exciton functions can be found from diagonalization of the
Hamiltonian
\begin{eqnarray}
\hat H_{2s/2p_y}= \left(
\begin{array}{cc}{\cal E}_{2s} & 3 e E_{y}a_\mathrm{B} \\
3 e E_{y}a_B & {\cal E}_{2p_y}
\end{array} \right), \label{eq:S1}
\end{eqnarray}
where $a_\mathrm{B}$ is the exciton Bohr radius, ${\cal
E}_{2s}={\cal E}_{2s_\mathrm{T}}$ is the zero-field energy of the
$2s$ transversal exciton, and  ${\cal E}_{2p_y}$ is that of the
$2p_y$ exciton state. The eigenenergies are
\begin{eqnarray}
{\cal E}_{2s_\mathrm{T}/2p_y}^\pm=\frac{1}{2}\left({\cal
E}_{2s_\mathrm{T}}+{\cal E}_{2p_y} \pm \sqrt{({\cal E}_{2p_y}-{\cal
E}_{2s_\mathrm{T}})^2+36(eE_{y}a_\mathrm{B})^2} \right),
\label{eq:S2}
\end{eqnarray}
and the resulting wave functions can be written as
\begin{eqnarray}
\Psi_{2s_\mathrm{T}/2p_y}=C_{2s_\mathrm{T}}(E_y)\Psi_{2s}+C_{2p_y}(E_y)\Psi_{2p_y} \label{eq:S3}
\end{eqnarray}
with
\begin{eqnarray}
C_{2s_{\mathrm{T}}}(E_y) = \frac{{\cal E}_{2p_y}-{\cal
E}_{2s_\mathrm{T}/2p_y}^\pm} {\sqrt{(3eE_{y}a_\mathrm{B})^2+({\cal
E}_{2p_y}-{\cal E}_{2s_\mathrm{T}/2p_y}^\pm)^2}} \label{eq:S4}
\end{eqnarray}
\begin{eqnarray}
C_{2p_y}(E_y) = -\frac{3eE_{y}a_\mathrm{B}}
{\sqrt{(3eE_{y}a_\mathrm{B})^2+({\cal E}_{2p_y}-{\cal
E}_{2s_\mathrm{T}/2p_y}^\pm)^2}} . \label{eq:S5}
\end{eqnarray}
In the considered geometry these exciton states are transversal
excitons. Due to interaction with the light field two transverse
polariton branches for each of the $2s/2p_y$ mixed states are
formed. The energies of the lower polariton branches (LPB) are given
by Eq.~\eqref{eq:S2}. To find the energies of the upper polariton
branches one has to consider the interaction of the mixed excitons
with photons and their corresponding contribution to the dielectric
function. However, when the longitudinal-transverse splitting,
$\Delta_{\mathrm{LT}}^{2s}$, is much smaller than all characteristic
energies, the results can be approximated by considering the direct
interaction between the $2p_y$ exciton and the upper $2s$ polariton
branch. For this, one has to use the ${\cal E}_{2s}={\cal
E}_{2s_\mathrm{L}}$ in the Hamiltonian \eqref{eq:S1}. The resulting energies
of the upper polariton branches (UPB) are approximately given by
\begin{eqnarray}
{\cal E}_{2s_\mathrm{L}/2p_y}^\pm  \approx \frac{1}{2}\left({\cal
E}_{2s_\mathrm{L}}+{\cal E}_{2p_y} \pm \sqrt{({\cal E}_{2p_y}-{\cal
E}_{2s_\mathrm{L}})^2+36(eE_{y}a_\mathrm{B})^2} \right) , \label{eq:S6}
\end{eqnarray}
where ${\cal E}_{2s_\mathrm{L}}={\cal
E}_{2s_\mathrm{T}}+\Delta_{\mathrm{LT}}^{2s}$ is the energy of the
$2s$-longitudinal exciton and the upper polariton in zero electric
field. The resulting wave functions can be found using
Eqs.~\eqref{eq:S2}-\eqref{eq:S5} after replacing the energy ${\cal
E}_{2s_\mathrm{T}}$ with ${\cal E}_{2s_\mathrm{L}}$.

\subsection{$1s$ excitons in external magnetic field perpendicular to hexagonal $z$-axis.}

The spin states of the $A$ and $C$  $1s$ excitons are formed from
the conduction band of $\Gamma_7$ symmetry and the valence band of
$\Gamma_7$ symmetry and thus can be of $\Gamma_5$, $\Gamma_1$ or
$\Gamma_2$ symmetry, split from each other by the electron-hole
exchange interaction. The external magnetic field $B_x$ mixes the
$\Gamma_{5y}$ and $\Gamma_1$ spin states. The resulting SHG active
states can be found from the Hamiltonian
\begin{eqnarray}
\hat H_{\Gamma_{5y}/\Gamma_1}= \left(
\begin{array}{cc}
{\cal E}_{\Gamma_1} & \mu_\mathrm{B} g_\mathrm{exc} B_x/2 \\
\mu_\mathrm{B} g_\mathrm{exc} B_x/2  & {\cal E}_{\Gamma_5}
\end{array} \right), \label{eq:S7}
\end{eqnarray}
where  ${\cal E}_{\Gamma_1}$  and  ${\cal E}_{\Gamma_5}$ are the
zero-field energies of the corresponding states, $g_{\mathrm{exc}} =
( g_\mathrm{h}^\bot-g_\mathrm{e}^\bot)$  is the effective $1s$
$g$-factor for $\textbf{B} \bot z$. In ZnO, for the $A$ exciton
$g_\mathrm{e}^\bot \approx 1.95$ and $g_\mathrm{h}^\bot\approx 0$
\cite{Lambrecht}. The resulting energies of the $1s$(\textit{A})-states in
magnetic field $B_x$ are given by
\begin{eqnarray}
{\cal E}_{\Gamma_{5y}/\Gamma_1}^\pm = \frac{1}{2}\left({\cal
E}_{\Gamma_1} + {\cal E}_{\Gamma_5} \pm
\sqrt{\Delta_{15}^2+(\mu_\mathrm{B} g_{\mathrm{exc}} B_x)^2} \right) ,  \label{eq:S8}
\end{eqnarray}
where $\Delta_{15}=|{\cal E}_{\Gamma_{5}}-{\cal E}_{\Gamma_1}|$ is
the exchange splitting. The LPB is described exactly by
Eq.~\eqref{eq:S8} with energy ${\cal E}_{\Gamma_5}={\cal
E}_{1s_{\mathrm{T}}}$, the UPB is described approximately by
Eq.~\eqref{eq:S8} with energy ${\cal E}_{\Gamma_5}={\cal
E}_{1s_{\mathrm{L}}}$. The resulting wave functions are given by
\begin{eqnarray}
\Psi_{\Gamma_5/\Gamma_1}=C_{\Gamma_5}(B_x)\Psi_{\Gamma_5}+C_{\Gamma_1}(B_x)\Psi_{\Gamma_1}\label{eq:S9}
\end{eqnarray}
with
\begin{eqnarray}
C_{\Gamma_5}(B_x) = \frac{2({\cal E}_{\Gamma_5}-{\cal
E}_{\Gamma_{5y}/\Gamma_1}^\pm)} {\sqrt{(\mu_\mathrm{B}
g_{\mathrm{exc}} B_x)^2+4({\cal E}_{\Gamma_5}-{\cal
E}_{\Gamma_{5y}/\Gamma_1}^\pm)^2}} \label{eq:S10}
\end{eqnarray}
\begin{eqnarray}
C_{\Gamma_1}(B_x) = -\frac{\mu_\mathrm{B} g_{\mathrm{exc}} B_x}
{\sqrt{(\mu_\mathrm{B} g_{\mathrm{exc}} B_x)^2+4({\cal
E}_{\Gamma_5}-{\cal E}_{\Gamma_{5y}/\Gamma_1}^\pm)^2}}.\label{eq:S11}
\end{eqnarray}

\subsection{$2s$ and $2p$ exciton states in magnetic field perpendicular to hexagonal $z$-axis.}

Similar to the effect of the external electric field $E_y$
considered above, the effective electric field $E_{\mathrm{eff}} =
\frac{\hbar}{M_{\text{exc}}} k_\mathrm{exc} B_x$, that originates
from the magneto-Stark effect [see Eq.~\eqref{Magneto-stark}], mixes
the $2s$ and $2p_y$ excitons of opposite parity. At the same time,
the Zeeman orbital effect mixes the $2p_z$ and $2p_y$ states of the
same parity. The resulting energies ${\cal E}_{2s/2p_z/2p_y}^{i}$
($i=1,2,3$ label eigenvalues) of the mixed $2s/2p_z/2p_y$ polariton
branches can be found as the eigenenergies of the Hamiltonian
\begin{eqnarray}
\hat H^\pm_{2s/2p_z/2p_y}= \left(
\begin{array}{ccc}
{\cal E}_{2s}^\pm(B_x) &0 & 3e E_{\mathrm{eff}}a_\mathrm{B} \\
 0& {\cal E}_{2p_z}^\pm(B_x)  & ig_{\mathrm{orb}}\mu_\mathrm{B} B_x\\
3 e E_{\mathrm{eff}}a_\mathrm{B}&-i g_{\mathrm{orb}}\mu_\mathrm{B}
B_x&{\cal E}_{2p_y}^\pm(B_x)
\end{array} \right). \label{eq:S12}
\end{eqnarray}
Here
\begin{eqnarray}
{\cal E}_{2s}^\pm(B_x)= E_{2s}+ 14 C_d B^2_x \pm  \mu_\mathrm{B} g_\mathrm{e}^\bot/2 \, , \\
{\cal E}_{2p_z}^\pm(B_x)={\cal E}_{2p_z}+ 12C_dB^2_x \pm \mu_\mathrm{B} g_\mathrm{e}^\bot/2 \, ,\\
{\cal E}_{2p_y}^\pm(B_x)={\cal E}_{2p_y}+ 12 C_dB^2_x \pm \mu_\mathrm{B} g_\mathrm{e}^\bot/2 \, ,
\end{eqnarray}
where $C_d$ describes the diamagnetic shift, and $g_{\mathrm{orb}}$
is the orbital $g$-factor. Expressions for $C_d$ and
$g_{\mathrm{orb}}$ can be found, for example, in
Ref.~\cite{Wheeler}. We assume the electron-hole short range
exchange splitting to be zero for all 2$s$ and 2$p$ states and
$g_h^\bot=0$, so that all states are additionally two times
degenerate with respect to the hole spin projection.

Accounting for the different spin states and polariton branches one
has to deal with $24$ mixed polariton states of the $A$ exciton and
$24$ mixed polariton states of the $B$ exciton. However, the
perturbations in Hamiltonian \eqref{eq:S12} mix only the states
belonging to the same spin states and the same polariton branches.
Therefore, in fact one has to consider only the
magnetic-field-induced mixing of the envelopes. Then the energies
${\cal E}^i_{2s_\mathrm{T}/2p_z/2p_y}$ of the LPB mixed states and
${\cal E}^i_{2s_\mathrm{L}/2p_z/2p_y}$ of the UPB mixed states can
be calculated as roots of the following equation:
\begin{widetext}
\begin{eqnarray}
({\cal E}^i-{\cal E}_{2s}^\pm)({\cal E}^i-{\cal E}_{2p_z}^\pm)
({\cal E}^i-{\cal
E}_{2p_y}^\pm)-(3eE_{\mathrm{eff}}a_\mathrm{B})^2({\cal E}^i-{\cal
E}_{2p_z}^\pm)-(g_{\mathrm{orb}}\mu_\mathrm{B} B_x)^2({\cal
E}^i-{\cal E}_{2s}^\pm)=0, \label{eq:S13}
\end{eqnarray}
with ${\cal E}_{2s}={\cal E}_{2s_\mathrm{T}}$ for the LPB and ${\cal
E}_{2s}={\cal E}_{2s_\mathrm{L}}$ for the UPB.

The resulting wave functions for the $2s/2p_z/2p_y$ mixed states
containall three components
\begin{eqnarray}
\Psi^i_{2s/2p_z/2p_y}=C^i_{2s}(B_x)\Psi_{2s}+C^i_{2p_z}(B_x)\Psi_{2p_z}+C^i_{2p_y}(B_x)\Psi_{2p_y}\label{eq:S14}
\end{eqnarray}
with the coefficients
\begin{eqnarray}
C^i_{2s}=\frac{3 e E_{\mathrm{eff}}a_\mathrm{B}({\cal E}^i-{\cal
E}_{2p_z}^\pm)} {\sqrt{({\cal E}^i-{\cal E}_{2s}^\pm)^2({\cal
E}^i-{\cal E}_{2p_z}^\pm)^2+(3 e
E_{\mathrm{eff}}a_\mathrm{B})^2({\cal E}^i-{\cal
E}_{2p_z}^\pm)^2+(g_{\mathrm{orb}}\mu_\mathrm{B} B_x)^2({\cal
E}^i-{\cal E}_{2s}^\pm)^2}} \, , \label{eq:S15}
\\
C^i_{2p_z} =\frac{ig_{\mathrm{orb}}\mu_\mathrm{B}
B_x({\cal E}^i-{\cal E}_{2s}^\pm)}{\sqrt{({\cal E}^i-{\cal
E}_{2s}^\pm)^2({\cal E}^i-{\cal E}_{2p_z}^\pm)^2+(3 e
E_{\mathrm{eff}}a_\mathrm{B})^2({\cal E}^i-{\cal
E}_{2p_z}^\pm)^2+(g_{\mathrm{orb}}\mu_\mathrm{B} B_x)^2({\cal
E}^i-{\cal E}_{2s}^\pm)^2}} \, , \label{eq:S16}
\\
C^i_{2p_y}=\frac{({\cal E}^i-{\cal E}_{2s}^\pm)({\cal E}^i-{\cal
E}_{2p_z}^\pm)}{\sqrt{({\cal E}^i-{\cal E}_{2s}^\pm)^2({\cal
E}^i-{\cal E}_{2p_z}^\pm)^2+(3 e
E_{\mathrm{eff}}a_\mathrm{B})^2({\cal E}^i-{\cal
E}_{2p_z}^\pm)^2+(g_{\mathrm{orb}}\mu_\mathrm{B} B_x)^2({\cal
E}^i-{\cal E}_{2s}^\pm)^2}} \, . \label{eq:S17}
\end{eqnarray}
\end{widetext}
It is worth to note, that the Hamiltonian \eqref{eq:S12} allows one
to take into account the effects of both the external magnetic field
$B_x$ and the external electric field $E_y$. For that purpose,
$E_\mathrm{eff}$ should be replaced with $E_\mathrm{eff} \pm E_y$,
where the choice of the sign depends on the direction of the applied
electric field.

The magnetic field $B_x$ affects the $2p_x$ exciton state only via
the spin Zeeman effect and the diamagnetic shift, but does not mix
it with the other $2p$ or $2s$ states. Its energy is given by
\begin{eqnarray}
{\cal E}_{2p_x}^\pm(B_x)={\cal E}_{2p_x}+ 6 C_dB^2_x \pm \mu_\mathrm{B}
g_\mathrm{e}^\bot/2 \, .  \label{eq:S18}
\end{eqnarray}
It is important to note, however, that the degeneracy of the $2p_x$
and $2p_y$ states is lifted by the magnetic field and the hexagonal
symmetry is broken, i.e., $\chi_{yyy}=\chi_{yxx}+2\chi_{xxy}$ is
violated for these states.


\begin{thebibliography}{99}
\bibitem{comment} Previous publications \onlinecite{Fiebig2, Wrzesinski} point out, that $1s$ excitons of the $A$- and $B$-series cannot be treated separately. Considering their coupling leads to the formation of three polariton branches, whereas the middle polariton branch is determined by $1s_L(A,\Gamma_5)$ for small $k$ and $1s_T(B, \Gamma_5)$ for large $k$, instead of two polariton branches for each series.
\end{thebibliography}
\end{document}